\documentclass[a4paper,11pt]{article}
\usepackage{jheppub}
\usepackage{lineno,color,slashed,changes}

\title{\boldmath Multi-component dark matter and Galactic 511 keV $\gamma$-ray emission}

\author[a,c]{Sarif Khan,}
\author[b,*]{Jinsu Kim,}
\author[c]{Jongkuk Kim,}
\author[d]{and Pyungwon Ko}
\note[*]{Corresponding author.}

\affiliation[a]{
    Institut f\"{u}r Theoretische Physik, 
    Georg-August-Universit\"{a}t G\"{o}ttingen,\\
    Friedrich-Hund-Platz 1,
    37077 G\"{o}ttingen, Germany
}
\affiliation[b]{
    School of Physics Science and Engineering, 
    Tongji University,\\
    Shanghai 200092, China
}
\affiliation[c]{
    Department of Physics,
    Chung-Ang University,\\
    Seoul 06974, Korea
}
\affiliation[d]{
    School of Physics,
    Korea Institute for Advanced Study,\\
    85 Hoegi-ro, Seoul 02455, Republic of Korea
}

\emailAdd{sarif.khan@uni-goettingen.de}
\emailAdd{kimjinsu@tongji.edu.cn}
\emailAdd{jongkukkim@cau.ac.kr}
\emailAdd{pko@kias.re.kr}

\preprint{KIAS-P24001}

\abstract{
  We study multi-component dark matter scenarios and the Galactic 511 keV $\gamma$-ray emission line signal in the framework of a local, dark $U(1)_D$ extension of the Standard Model. A light vector dark matter particle associated with the dark $U(1)_D$ may decay and annihilate to electron-positron pairs. The produced positrons may in turn form positroniums that subsequently annihilate to two photons, accounting for the observed line signal of the Galactic 511 keV $\gamma$-ray emission. Three scenarios are investigated. First, we consider the minimal $U(1)_D$ extension where a dark gauge boson and a dark Higgs boson are newly introduced to the particle content. As a second scenario, we consider WIMP-type dark matter with the introduction of an extra dark fermion which, in addition to the dark gauge boson, may contribute to the dark matter relic abundance. It is thus a multi-component dark matter scenario with a UV-complete dark $U(1)_D$ symmetry. In particular, the vector dark matter may account for a small fraction of the total dark matter relic abundance. Finally, we consider the scenario where the dark matter particles are of the FIMP-type. In this case, both the light vector and fermion dark matter particles may be produced via the freeze-in and super-WIMP mechanisms. Considering theoretical and observational constraints, we explore the allowed parameter space where the Galactic 511 keV $\gamma$-ray line signal and the dark matter relic can both be explained. We also discuss possible observational signatures.
}

\begin{document}
\maketitle
\flushbottom

\section{Introduction}
\label{sec:intro}
The Standard Model (SM) successfully describes the fundamental particles and their interactions via local gauge theory based on $SU(3)_C \times SU(2)_L \times U(1)_Y$. Nevertheless, the existence of dark matter (DM) has been confirmed by a number of observations such as rotation curves, gravitational lensing, bullet clusters, large-scale structure formation, and the shape of the cosmic microwave background (CMB) anisotropy power spectrum, all of which are only through gravitational interactions. So far, no other interactions of DM are known, and its physical identity remains a mystery. 
The CMB observation from the Planck satellite indicates that the total DM relic density is given by \cite{Planck:2018vyg}
\begin{align}
  \Omega_{\rm DM} h^2 = 0.1200  \pm 0.0012.
\end{align}

There have been numerous studies on the nature of DM; see, {\it e.g.}, Refs.~\cite{Profumo:2019ujg,Tuominen:2021wrl,Arbey:2021gdg} for a recent review. One of the most promising DM candidates is the so-called Weakly Interacting Massive Particle (WIMP). The total WIMP-type DM annihilation cross section required to obtain the correct relic density is \cite{Steigman:2012nb}
\begin{align}
  \langle \sigma v \rangle_{\rm tot} \simeq 3\times 10^{-26}
  \, {\rm cm^3/s}
  \,.
\end{align}
However, the null result from the direct detection experiments implies that DM might interact with the SM particles very weakly. If DM feebly interacts with the SM particles in the early Universe, it may have never entered the thermal equilibrium. In this case, the DM may be produced via the freeze-in mechanism \cite{McDonald:2001vt} rather than the standard freeze-out mechanism. The DM of this type is dubbed Feebly Interacting Massive Particle (FIMP) \cite{Hall:2009bx}; see also Ref.~\cite{Bernal:2017kxu} for a review on FIMP DM.
Another interesting DM production mechanism is known as the super-WIMP mechanism \cite{Covi:1999ty,Feng:2003uy,Feng:2003xh}, where a meta-stable, frozen-out particle eventually decays to the DM.

The DM is not only capable of explaining the missing energy budget of our Universe, but it may also account for the 511 keV $\gamma$-ray emission line signal at the Galactic Centre (GC) which was first observed more than 50 years ago \cite{Johnson:1972apj}. The consistent picture has been observed by analyses of data from CGRO/OSSE \cite{Purcell:1997apj} and INTEGRAL/SPI \cite{Bouchet:2010dj,Bouchet:2011fn,Siegert:2019tus} as well. While the physical origin of this excess is yet to be resolved, DM could be a good candidate; for a review, see, {\it e.g.}, Refs.~\cite{Prantzos:2010wi,Leane:2022bfm,Siegert:2023wus}.
Various explanations for the 511 keV $\gamma$-ray signal with DM have been proposed \cite{Boehm:2003bt,Hooper:2004qf,Picciotto:2004rp,Gunion:2005rw,Takahashi:2005kp,Finkbeiner:2007kk,Huh:2007zw,Pospelov:2007xh,Cembranos:2008bw,Arkani-Hamed:2008hhe,Khlopov:2008ki,Khlopov:2009hi,Chen:2009av,Chen:2009dm,Finkbeiner:2009mi,Khlopov:2010pq,Cline:2010kv,Cline:2012yx,Boubekeur:2012eq,Cudell:2014jba,Cudell:2014wca,Farzan:2017hol,Jia:2017iyc,Cai:2020fnq,Ema:2020fit,Lin:2022mqe,Sheng:2023iup,Feng:2024nkh}. The 511 keV signal has also been explained using the electric field property of a black hole in Ref.~\cite{Dolgov:2023sst}.
The line signal of 511 keV photon may come from the decay process of positronium, which would have been formed by interactions of positrons and electrons. The DM could provide abundant positrons via DM decay and/or pair-annihilation processes. To explain the Galactic 511 keV $\gamma$-ray line signal, the injection energy of the produced positrons is required to be less than $\sim$3 MeV \cite{Beacom:2005qv}. Furthermore, if annihilation of DM to an electron-positron pair were to be considered, the annihilation cross section is required to be
\begin{align}
  (\sigma v)_{{\rm DM}\,{\rm DM} \to e^+e^-} \simeq
  10^{-30} \left( \frac{m_{\rm DM}}{\rm MeV} \right)^2
  \,{\rm cm^3/s}
  \,,
\end{align}
for the Navarro–Frenk–White (NFW) profile.
Such a small cross section is much smaller than the cross section of $10^{-26}\;{\rm cm^3/s}$ required for a single-component WIMP-type, thermal freeze-out DM. It is thus challenging to explain the 511 keV signal with the standard DM picture. For more implications of recent data analyses, readers may refer to Refs.~\cite{Ascasibar:2005rw,Vincent:2012an,Wilkinson:2016gsy,Cappiello:2023qwl,DelaTorreLuque:2023cef}.

In this work, we aim to explain the correct DM relic abundance and the Galactic 511 keV $\gamma$-ray emission line signal in the framework of a local, dark $U(1)_D$ extension of the SM. The dark gauge boson associated with the dark $U(1)_D$ may become responsible for the 511 keV signal. After discussing the minimal setup of the $U(1)_D$ extension, we move on to include a dark fermion which can be another good DM candidate, realising a multi-component DM scenario. We investigate both WIMP-type and FIMP-type DM scenarios and explore viable parameter spaces for the correct DM relic density and the 511 keV signal. Various theoretical and observational constraints are also considered, and we shall explicitly identify the allowed parameter region.

The paper is organised as follows. In Sec.~\ref{sec:minimal}, we first consider the minimal $U(1)_D$ extension where a dark gauge boson and a dark Higgs boson are the only newly introduced particles. We discuss how a light, dark gauge boson may source the Galactic 511 keV $\gamma$-ray signal.
In Sec.~\ref{sec:2Comp}, we introduce a dark fermion to the minimal scenario and investigate two multi-component DM scenarios in connection with the 511 keV signal, depending on the $U(1)_D$ charge of the dark fermion. A thorough numerical computation is performed to identify the viable parameter space for both the correct DM relic abundance and the 511 keV signal.
We conclude in Sec.~\ref{sec:conc}.
Appendix~\ref{apdx:decaywidths} summarises decay width expressions for various processes that are relevant in the current study.

\section{The Minimal Scenario}
\label{sec:minimal}
We first consider the minimal setup of a $U(1)_D$ extension of the SM where a dark Higgs field, $\phi_D$, and a dark gauge boson, $\hat{X}$, associated with the $U(1)_D$ are the only extra fields.
The Lagrangian is given by
\begin{align}
    \mathcal{L}_{\rm M} = \mathcal{L}_{\overline{\rm SM}} 
    - \frac{1}{4}\hat{X}_{\mu\nu}\hat{X}^{\mu\nu}
    -\frac{\sin\epsilon}{2}\hat{X}_{\mu\nu}\hat{B}^{\mu\nu}
    + |D\phi_D|^2 + |DH|^2 
    - V(\phi_D,H)
    \,,
    \label{eqn:LagMinimal}
\end{align}
where $\mathcal{L}_{\overline{\rm SM}}$ is the SM Lagrangian excluding the SM Higgs sector except for the Yukawa interaction terms, $\hat{X}_{\mu\nu}$ is the field-strength of the dark gauge boson $\hat{X}$, $\hat{B}_{\mu\nu}$ is the field-strength of the hypercharge gauge boson, $\epsilon$ parametrises the $U(1)_Y$--$U(1)_D$ gauge kinetic mixing, $H$ is the SM Higgs field, $\phi_D$ is the dark Higgs field, and the covariant derivative for the dark sector is $D_\mu\phi_D = \partial_\mu\phi_D - i g_D \hat{X}_\mu n_{\phi_D} \phi_D$ with $g_D$ being the dark gauge coupling and $n_{\phi_D}$ the $U(1)_D$ charge of the dark Higgs field. Without loss of generality, we set $n_{\phi_D}=1$ in the current work.
Finally, the potential $V(\phi_D,H)$ is given by
\begin{align}
    V(\phi_D,H) = -\mu_D^2|\phi_D|^2
    +\lambda_D|\phi_D|^4
    -\mu_H^2|H|^2
    +\lambda_H|H|^4
    +\lambda_{HD}|\phi_D|^2|H|^2\,.
\end{align}

The gauge kinetic mixing term can be removed via the gauge boson mass matrix diagonalisation process \cite{Babu:1997st}. Closely following the procedures and notations\footnote{Instead of $Z'$ used in Ref.~\cite{Choi:2021yps}, we use $Z_D$.} outlined in Ref.~\cite{Choi:2021yps}, we define $\tilde{B}_\mu$ and $\tilde{X}_\mu$ as
\begin{align}
    \left(\begin{array}{c}
        \hat{B}_\mu \\
        \hat{X}_\mu
    \end{array}\right) =
    \left(\begin{array}{cc}
        1 & -\tan\epsilon \\
        0 & 1/\cos\epsilon
    \end{array}\right)
    \left(\begin{array}{c}
        \tilde{B}_\mu \\
        \tilde{X}_\mu
    \end{array}\right)
    \,,
\end{align}
and $\hat{W}_\mu = \tilde{W}_\mu$.
The diagonalisation of the mass matrix for the gauge fields $\tilde{B}_\mu$, $\tilde{X}_\mu$, and $\tilde{W}^3_\mu$ can then be done through
\begin{align}
    \left(\begin{array}{c}
        \tilde{B}_\mu \\
        \tilde{W}^3_\mu \\
        \tilde{X}_\mu 
    \end{array}\right) =
    \left(\begin{array}{ccc}
        \cos\theta_w & -\sin\theta_w \cos\zeta & \sin\theta_w \sin\zeta \\
        \sin\theta_w & \cos\theta_w \cos\zeta & -\cos\theta_w \sin\zeta \\
        0 & \sin\zeta & \cos\zeta 
    \end{array}\right)
    \left(\begin{array}{c}
        A_\mu \\
        Z_\mu \\
        Z_{D\mu}
    \end{array}\right)
    \,,
\end{align}
where $\theta_w$ is the Weinberg angle, and the angle $\zeta$ is defined via
\begin{align}
    \tan 2\zeta \equiv -\frac{m^2_{\hat{Z}} \sin\theta_w\sin2\epsilon}{m^2_{\hat{X}} - m^2_{\hat{Z}} (\cos^2\epsilon - \sin^2\epsilon\sin^2\theta_w)}\,,
\end{align}
with $m_{\hat{Z}} = \sqrt{g_1^2 + g_2^2}v_H/2$, $m_{\hat{X}} = g_D v_D$, and $v_H$ ($v_D$) being the vacuum expectation value (VEV) of the SM (dark) Higgs field.
One may obtain the masses of the $Z$ and $Z_D$ as
\begin{align}
    m_Z^2 = m_{\hat{Z}}^2\left(1 + \sin\theta_w\tan\zeta\tan\epsilon\right)
    \,,\qquad
    m_{Z_D}^2 = \frac{m_{\hat{X}}^2}{\cos^2\epsilon(1 + \sin\theta_w\tan\zeta\tan\epsilon)}
    \,,
    \label{eqn:ZZDmasses}
\end{align}
and the relations between the original fields and the mass eigenstates as
\begin{align}
    \hat{B}_\mu &=
    \cos\theta_w A_\mu
    - \left(\tan\epsilon\sin\zeta+\sin\theta_w\cos\zeta\right)Z_\mu
    - \left(\tan\epsilon\cos\zeta-\sin\theta_w\sin\zeta\right)Z_{D\mu}
    \,,\nonumber\\
    \hat{X}_\mu &=
    \frac{\sin\zeta}{\cos\epsilon}Z_\mu 
    + \frac{\cos\zeta}{\cos\epsilon}Z_{D\mu}
    \,,\nonumber\\
    \hat{W}^3_\mu &=
    \sin\theta_w A_\mu 
    + \cos\theta_w\cos\zeta Z_\mu 
    - \cos\theta_w\sin\zeta Z_{D\mu}
    \,.
\end{align}

The mass matrix for the SM and dark Higgs fields can also be diagonalised. Expressing the Higgs fields in the unitary gauge as
\begin{align}
    H = \frac{1}{\sqrt{2}}\left(\begin{array}{c}
        0 \\
        v_H + h
    \end{array}\right)
    \,,\qquad
    \phi_D = \frac{1}{\sqrt{2}}\left(
        v_D + \phi 
    \right)
    \,,
\end{align}
the mass matrix is given by
\begin{align}
    m_{h\phi}^2 = \left(\begin{array}{cc}
        2\lambda_H v_H^2 & \lambda_{HD}v_Hv_D \\
        \lambda_{HD}v_Hv_D & 2\lambda_D v_D^2
    \end{array}\right)
    \,.
\end{align}
The mass eigenstates are then obtained as follows:
\begin{align}
    h_1 &= h \cos\theta + \phi \sin\theta
    \,,\nonumber\\
    h_2 &= \phi \cos\theta - h \sin\theta
    \,,
\end{align}
where the mixing angle $\theta$ is defined via
\begin{align}
    \tan2\theta = \frac{\lambda_{HD}v_Hv_D}{\lambda_H v_H^2 - \lambda_D v_D^2}
    \,,
\end{align}
while the mass eigenvalues are given by
\begin{align}
    m_{h_{1,2}}^2 =
    \lambda_H v_H^2 + \lambda_D v_D^2
    \mp 
    \sqrt{\left(\lambda_H v_H^2 - \lambda_D v_D^2\right)^2 + \lambda_{HD}^2v_H^2v_D^2}
    \,.
\end{align}

The minimal scenario has one DM candidate, namely the dark $U(1)_D$ gauge boson $Z_D$. Its DM phenomenology has been extensively studied by many authors; see, {\it e.g.}, Refs.~\cite{Bauer:2018egk,Fabbrichesi:2020wbt} for a recent review.
In order for $Z_D$ to play the role of DM, its lifetime is required to be longer than the age of the Universe. It gives an upper bound on the kinetic mixing parameter, $\epsilon \lesssim 10^{-20}$, for $m_{Z_D}=\mathcal{O}(10)$ GeV \cite{Costa:2022lpy}.\footnote{
The $\gamma$-ray observation puts a much stronger bound of $\epsilon \lesssim 10^{-26}$ \cite{Fermi-LAT:2015kyq}, assuming that $Z_D$ constitutes 100\% of DM.
}
Let us examine implications of the minimal scenario on the 511 keV emission line from the GC.
Taking the mass of the dark gauge boson to be of the MeV order, $Z_D$ may decay into a pair of electron and positron, $Z_D \rightarrow e^-e^+$. It also has an annihilation channel to an electron-position pair, $Z_D Z_D \rightarrow e^-e^+$. The positron production rate is then given, assuming that $Z_D$ constitutes the total DM relic density today, by
\begin{align}
    \dot{n}_{e^+} &=
    \frac{1}{m_{Z_D}}\Gamma_{Z_D \rightarrow e^-e^+}\rho_{\rm DM}
    + \frac{1}{2m_{Z_D}^2}(\sigma v)_{Z_DZ_D\rightarrow e^-e^+}\rho_{\rm DM}^2\,,
\end{align}
where $m_{Z_D}$ is the mass of the DM given in Eq.~\eqref{eqn:ZZDmasses}, $\Gamma$ is the decay width, $(\sigma v)$ is the annihilation cross section times velocity in the non-relativistic limit, and $\rho_{\rm DM}$ is the Galactic DM profile which we choose to be the NFW profile \cite{Navarro:1995iw},
\begin{align}
    \rho_{\rm DM} = \frac{\rho_s}{r/r_s(1+r/r_s)^2}\,,
\end{align}
with $\rho_s \approx 0.184 \; {\rm GeV}/{\rm cm}^3$ and $r_s \approx 24.42\; {\rm kpc}$ \cite{Cirelli:2010xx}.
The produced positrons would form positronium atoms via interactions with electrons of neutral hydrogen atoms \cite{Beacom:2005qv}.\footnote{
The formation of positronium becomes viable once the produced positrons have enough energy so that the threshold of 6.8 eV can be overcome. It is, however, possible, through free electrons in ionised gas, to form positronium when the energy of the produced positrons is insufficient to overcome the threshold; it is through free electrons in ionised gas; see, {\it e.g.}, Refs.~\cite{Prantzos:2010wi,Sheng:2023iup}.
}
A quarter of the positronium then annihilates to two photons with 511 keV. Therefore, assuming that all the produced positrons form positronium and that the annihilation happens immediately as they are produced, we obtain $\dot{n}_\gamma = \dot{n}_{e^+}/2$.
The (differential) photon flux can be computed by integrating along the line of sight (l.o.s.),
\begin{align}
    \frac{d\Phi^{511}_\gamma}{d\Omega} = \frac{1}{4\pi}\int_{\rm l.o.s.}\dot{n}_\gamma \, ds
    \,.
\end{align}
Thus, we find, taking the decay rate and the thermally-averaged cross section to be constant along the l.o.s., that 
\begin{align}
    \frac{d\Phi^{511}_\gamma}{d\Omega} &=
    \frac{\Gamma_{Z_D\rightarrow e^-e^+}}{8\pi m_{Z_D}}
    \int_{\rm l.o.s.}\rho_{\rm DM}\,ds
    +\frac{(\sigma v)_{Z_DZ_D\rightarrow e^-e^+}}{16\pi m_{Z_D}^2}
    \int_{\rm l.o.s.}\rho_{\rm DM}^2\,ds
    \nonumber\\&=
    \frac{\Gamma_{Z_D\rightarrow e^-e^+}}{8\pi m_{Z_D}}R_\odot\rho_\odot J_{\rm decay}(\psi)
    +\frac{(\sigma v)_{Z_DZ_D\rightarrow e^-e^+}}{16\pi m_{Z_D}^2}R_\odot\rho_\odot^2 J_{\rm ann}(\psi)
    \,,
    \label{eqn:diffPhotonFlux}
\end{align}
where
\begin{align}
    J_{\rm decay} \equiv 
    \int_{\rm l.o.s.}
    \frac{\rho_{\rm DM}}{\rho_\odot}
    \frac{ds}{R_\odot}
    \,,\qquad
    J_{\rm ann} \equiv 
    \int_{\rm l.o.s.}
    \left(\frac{\rho_{\rm DM}}{\rho_\odot}\right)^2
    \frac{ds}{R_\odot}
    \,,
\end{align}
with $R_\odot \approx 8.33\;{\rm kpc}$ and $\rho_\odot \approx 0.3\;{\rm GeV}/{\rm cm}^3$ \cite{Cirelli:2010xx}.
The integrals over the l.o.s. can be performed by noting that $r=\sqrt{s^2+R_\odot^2-2sR_\odot\cos \psi}$, where $\cos\psi\equiv\cos b\cos l$ with $b$ and $l$ being respectively the latitude and longitude in Galactic coordinates.
In order to compare with the observed photon flux, we integrate the differential photon flux \eqref{eqn:diffPhotonFlux} over $\psi<10.275^\circ$ \cite{Siegert:2015knp},
\begin{align}
    \Phi_\gamma^{511} = \int_{\psi<10.275^\circ}\frac{d\Phi_\gamma^{511}}{d\Omega}\,d\Omega \,.
    \label{eqn:PhotonFlux}
\end{align}
We note that
\begin{align}
    \int_{\psi<10.275^\circ} J_{\rm decay} \, d\Omega \approx 1.2
    \,,\qquad
    \int_{\psi<10.275^\circ} J_{\rm ann} \, d\Omega \approx 9.7
    \,.
\end{align}
We then compare the result with the observed value of $(1/2)\times (0.96\pm0.07) \times 10^{-3} \; {\rm cm}^{-2}{\rm sec}^{-1}$ \cite{Siegert:2015knp}.

\begin{figure}[t!]
    \centering
    \includegraphics[scale=0.3]{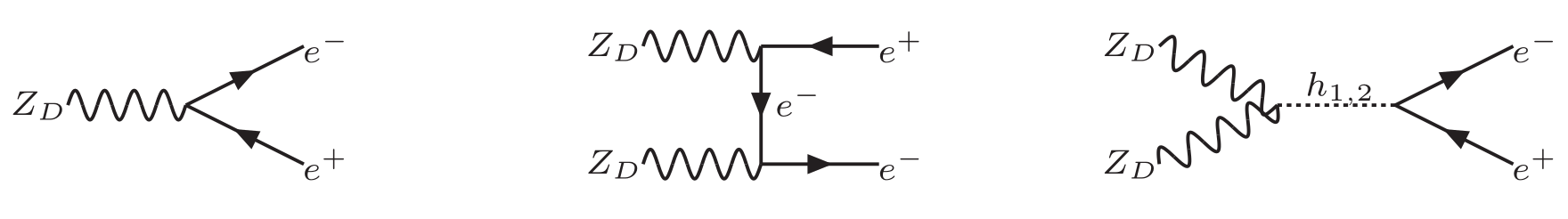}
    \caption{Feynman diagrams for $Z_D\rightarrow e^- e^+$ and $Z_DZ_D\rightarrow e^-e^+$ processes.}
    \label{fig:diagram_ee}
\end{figure}

The remaining ingredients are the particle physics-related quantities, namely the decay width $\Gamma_{Z_D\rightarrow e^-e^+}$ and the usual annihilation cross section times velocity in the non-relativistic limit $(\sigma v)_{Z_DZ_D\rightarrow e^-e^+}$.
The relevant Feynman diagrams are shown in Fig.~\ref{fig:diagram_ee}.
The decay rate is given by
\begin{align}
    \Gamma_{Z_D\rightarrow e^-e^+} &=
    \frac{m_{Z_D}e^2\cos^2\zeta}{12\pi\cos^2\theta_w\sin^2\theta_w}
    \sqrt{1-\frac{4m_e^2}{m_{Z_D}^2}}
    \nonumber\\
    &\qquad\qquad\times
    \left[
    V_{Z_De^-e^+}^2 + A_{Z_De^-e^+}^2
    +\frac{2m_e^2}{m_{Z_D}^2}\left(V_{Z_De^-e^+}^2-2A_{Z_De^-e^+}^2\right)
    \right]
    \,,
\end{align}
where
\begin{align}
    V_{Z_De^-e^+} &=
    \sin\theta_w\left(\tan\epsilon-\sin\theta_w\tan\zeta\right)-\frac{1}{4}\left(\sin\theta_w\tan\epsilon-\tan\zeta\right)
    \,,\nonumber\\
    A_{Z_De^-e^+} &=
    \frac{1}{4}\left(\sin\theta_w\tan\epsilon-\tan\zeta\right)
    \,,
\end{align}
while the usual annihilation cross section times velocity in the non-relativistic limit is 
\begin{align}
    (\sigma v)_{Z_D Z_D\rightarrow e^-e^+} &\approx  \frac{\cos^2\theta\sin^2\theta y_e^2 v_D^2 g_D^4 (m_{h_1}^2 - m_{h_2}^2)^2}{ 12\pi(m_{h_1}^2-4m_{Z_D}^2)^2(m_{h_2}^2-4m_{Z_D}^2)^2}\left(   1-\frac{m_e^2}{m_{Z_D}^2}  \right)^{3/2}\,,
\end{align}
where we have neglected the contribution coming from the second diagram of Fig.~\ref{fig:diagram_ee}, given that the gauge kinetic mixing angle is much smaller than the Higgs mixing angle.

As we discussed in the introduction, it is difficult to explain the Galactic 511 keV $\gamma$-ray line signal from the annihilation of MeV-scale $Z_D$ dark matter; the required thermally-averaged cross section is
\begin{align}
  \langle \sigma v \rangle_{Z_DZ_D\rightarrow e^-e^+} \simeq
  \frac{1}{f^2_{Z_{D}}} \times
  \left(
  \frac{m_{Z_D}}{{\rm MeV}}
  \right)^2 \times 10^{-30}\;{\rm cm^3/s}\,,
\end{align}
where the prefactor $f_{Z_D} = \Omega_{Z_D}h^2/0.12$ is the fraction of the vector DM to the total DM relic density. We see that it is in conflict with the cross section required for the correct relic density if $f_{Z_D} = 1$.
If we were to use the $Z_D$ decay to explain the 511 keV photon flux, the decay width, or equivalently, the lifetime, needs to satisfy
\begin{align}
  \tau_{Z_D} \simeq
f_{Z_{D}} \times  \left(\frac{{\rm MeV}}{m_{Z_D}}\right) \times 8 \times 10^{26}\;{\rm s}\,.
\end{align}
The $Z_D$ lifetime can be much longer than the age of the Universe. As is well known, it in turn requires that the gauge kinetic mixing angle be tiny. The remaining task would then be to answer whether or not the correct DM relic density could successfully be achieved in this case.
In the next section, we introduce an extra dark fermion to the minimal scenario, realising a multi-component DM scenario. As we shall see, the minimal scenario is in fact a limiting case of the multi-component DM scenario. We present the detailed analysis in the next section and discuss the limiting case of the minimal scenario there as well. We will see that the extra fermionic DM mainly annihilates to the vector DM, and the vector DM annihilates to the dark Higgs boson, if kinematically open. By tuning the vector DM fraction and its lifetime, we can explain the 511 keV line, and the rest of the DM can be accommodated by the fermionic DM, suitably choosing its $U(1)_D$ gauge charge. Therefore, we can meet the total amount of DM obtained by the Planck satellite.

\section{Multi-Component Scenarios}
\label{sec:2Comp}
Having discussed the minimal scenario, we explore in this section multi-component scenarios by introducing an extra dark fermion $\psi_D$ that is charged under the dark $U(1)_D$ gauge group. The DM candidates then become both $\psi_D$ and $Z_D$.
The 511 keV aspect in the presence of both the dark gauge boson and dark fermion has been investigated in, {\it e.g.}, Refs.~\cite{Huh:2007zw,Pospelov:2007mp}. We note, however, that they do not take into account the presence of a dark Higgs field and the Higgs-portal interaction. Furthermore, in Ref.~\cite{Huh:2007zw}, a massless dark gauge boson is considered, while in Ref.~\cite{Pospelov:2007mp}, the decay of $Z_D$ to $e^-e^+$ is not thoroughly investigated.
We consider in this work a more complete model, taking into account not only the presence of the dark Higgs field, but also the decay channel of the massive dark gauge boson $Z_D$.
Two scenarios may be assessed; one where the dark $U(1)_D$ charge of the dark fermion is not 1/2, and the other where the dark fermion has the dark charge of 1/2 in which case a Yukawa term is allowed. In the following subsections, we examine these two scenarios.

\subsection{$n_{\psi_D} \neq 1/2$ Case}
\label{subsec:GenCase}
Dark matter phenomenology in this case has been studied in Ref.~\cite{Khan:2023uii} which mainly focused on the GeV scale DM sector. In this work, we aim to explain the Galactic 511 keV emission line signal, and thus, we focus on the case where the dark gauge boson $Z_D$ is in the MeV scale.
The Lagrangian is given by
\begin{align}
    \mathcal{L}_{n_{\psi_D}\neq 1/2} = \mathcal{L}_{\rm M}
    + \bar{\psi}_D\left(i\gamma^\mu D_\mu - m_{\psi_D}\right)\psi_D
    \,,
    \label{eqn:LagGen}
\end{align}
where $\mathcal{L}_{\rm M}$ is the Lagrangian for the minimal scenario, given in Eq.~\eqref{eqn:LagMinimal}, and the covariant derivative is given by $D_\mu\psi_D = \partial_\mu\psi_D - i g_D n_{\psi_D} \hat{X}_\mu\psi_D$ with $n_{\psi_D}\neq 1/2$ being the dark $U(1)_D$ charge of the dark fermion.

In this consideration, we can realise a multi-component DM scenario depending on the mass of the additional gauge boson. The dark gauge boson $Z_D$ and the dark fermion $\psi_D$ become good DM candidates if we choose $m_{Z_D} < 2 m_{\psi_D}$.
The evolution of $Z_D$ and $\psi_D$ DM candidates is governed by the Boltzmann equations,
\begin{align}
    \frac{d Y_{i}}{d x_i} = - \frac{s(x_i) \langle \sigma v \rangle_i}{x_i H(x_i)} 
    \left[ Y^{2}_{i} - \left( Y^{\rm eq}_{i} \right)^2 \right]
    \,,\qquad 
    i = \{Z_D, \psi_D\}
    \,,
\end{align}
where $Y_i=n_i/s$ is the yield of the DM, $n_i$ is the number density, $s$ is the entropy density, $H$ is the Hubble parameter, $x_i = m_i/T$, and $\langle \sigma v \rangle_i$ is the relevant thermally-averaged cross section for the DM species $i$. The entropy density and the Hubble parameter can be expressed as
\begin{align}
    s(x_i) = \frac{2 \pi^2}{45} g_{*,s} m^3_{i} x_i^{-3}
    \,,\quad
    H(x_i) = \sqrt{\frac{g_* \pi^2}{90}} \frac{m^2_i}{M_{\rm P}} x_i^{-2}
    \,,
\end{align}
with $M_{\rm P}$ and $g_{*,s}$ ($g_*$) being, respectively, the reduced Planck mass and effective entropy (energy) degrees of freedom.
In the present work, we consider the vector DM in the MeV range and the fermionic DM in the GeV range, which makes them thermally decouple at different temperatures during Universe's evolution.
For the analytical estimate, we may evolve the Boltzmann equations independently. We have checked with {\tt micrOMEGAs}~\cite{Alguero:2023zol} that it is a valid approximation for the parameter space we explore in the current work.
In general, if the vector and fermionic DM masses are close, we would have $\psi_{D} \bar{\psi}_{D} \leftrightarrow Z_{D}Z_{D}$. 
Therefore, the abundances of $\psi_D$ and $Z_D$ are set by the processes $\psi_{D} \bar{\psi}_{D} \leftrightarrow Z_{D}Z_{D}$ and $Z_{D}Z_{D} \leftrightarrow h_{1,2}h_{1,2}$. As a result, at the onset of the $\psi_D$ freeze-out, $Z_D$ can still be in thermal equilibrium through the annihilation processes that involve the Higgses.
The $Z_D$ then goes out of equilibrium later for the current choice of the mass spectrum.
One may also expect, from the kinetic mixing between the gauge bosons, the process $Z_D \psi_D \leftrightarrow Z \psi_D$ that can change the $Z_D$ abundance. However, this process is ineffective in the current work because we are considering a small gauge kinetic mixing.

The thermally-averaged cross section can be expanded as $\langle \sigma v \rangle_i = a_i + 6b_i/x_i$. The coefficients $a_i$ and $b_i$ for the $Z_D$ and $\psi_D$ DM are obtained by considering $Z_D Z_D \rightarrow h_2 h_2$ for $i=Z_D$ and $\psi_D\bar{\psi}_D \rightarrow Z_D Z_D$ for $i=\psi_D$; they are given by
\begin{align}
    a_{Z_D} &= \frac{11 g_D^4 \sqrt{1 - r_1}}{288 \pi m_{Z_D}^2}
    \left[
    \frac{1-\frac{20}{11}r_1+\frac{15}{11}r_1^2-\frac{5}{11}r_1^3+\frac{1}{16}r_1^4}{(1-\frac{3}{4}r_1+\frac{1}{8}r_1^2)^2}
    \right]
    \,,\nonumber\\
    b_{Z_D} &= \frac{13 g_D^4 r_1 \sqrt{1 - r_1}}{1728 \pi m_{Z_D}^2}\left[
    \frac{1 - \frac{19}{13}r_1 + \frac{22}{13}r_1^2 - \frac{17}{13}r_1^3 + \frac{127}{208}r_1^4 - \frac{17}{104}r_1^5 + \frac{1}{52}r_1^6}{(1-\frac{1}{2}r_1)^4(1-\frac{1}{4}r_1)^3}
    \right]
    \,,\nonumber\\
    a_{\psi_D} &= \frac{g_D^4 n_{\psi_D}^4 \cos^4\zeta (1-r_2)^{3/2}}{16\pi m_{\psi_D}^2 \cos^4\epsilon \left(1-\frac{1}{2}r_2\right)^2}
    \,,\nonumber\\
    b_{\psi_D} &= \frac{g_D^4 n_{\psi_D}^4 \cos^4\zeta \sqrt{1-r_2}}{8\pi m_{\psi_D}^2 \cos^4\epsilon}\left[\frac{1-\frac{2}{3}r_2+\frac{1}{8}r_2^2+\frac{1}{6}r_2^3}{\left(1-\frac{1}{2}r_2\right)^4}\right]
    \,,
    \label{ab-psi-wd}
\end{align}
where $r_1 \equiv (m_{h_2}/m_{Z_D})^2$ and $r_2 \equiv (m_{Z_D}/m_{\psi_D})^2$.
We note that in obtaining the cross section for $Z_D Z_D \rightarrow h_2 h_2$, we have ignored the small contributions from $Z$ and $h_1$. Furthermore, we have taken leading order terms in both the Higgs and the gauge kinetic mixing angles.

Solving the Boltzmann equations, one may compute the yield $Y^0_i$ of DM particle $i$ at present as \cite{Gondolo:1990dk,Edsjo:1997bg}
\begin{align}
	Y^0_i = Y^f_i \left[
	1 + \frac{2\pi^2}{45} \sqrt{\frac{90}{\pi^2}} M_{\rm P} m_i Y^f_i
	\frac{g_{*,s}(x_i^f)}{\sqrt{g_*(x_i^f)}} \frac{a_i + 3b_i / x_i^f}{x_i^f}
	\right]^{-1}
	\,,\label{eqn:YieldToday}
\end{align}
with
\begin{align}
	Y^f_i = \frac{45}{2\pi^4} \sqrt{\frac{\pi}{8}} \frac{g_i}{g_{*,s}(x_i^f)}(x_i^f)^{3/2}e^{-x_i^f}
	\,.
\end{align}
Here, $g_i$ is the internal degree of freedom of the DM particle $i$, and $x_i^f$ is the freeze-out time at which
\begin{align}
n^{\rm eq}_i \langle \sigma v \rangle_i  \simeq H(x_i)\,\, {\rm at} \,\,x_i=x_i^f\,,     \label{equilibrium-condition}
\end{align}
where $n^{\rm eq}_i$ is equilibrium number density of the DM particle $i$.
We numerically solve this equation to obtain the freeze-out temperature $x^f_i$, taking into account the temperature dependence of the effective degrees of freedom $g_{*,s}$ and $g_*$.
The dark matter relic density can then be computed as
\begin{align}
	\Omega_{\rm DM} h^2 = 
	\Omega_{Z_D} h^2 + \Omega_{\psi_D} h^2 =
	\frac{s_0 h^2}{\rho_{\rm crit}}\Big(
	m_{Z_D}Y_{Z_D}^0 + m_{\psi_D}Y_{\psi_D}^0
	\Big)
	\,,
\end{align}
where $s_0 \approx 2891.2 \; {\rm cm}^{-3}$ is the entropy density today, $\rho_{\rm crit} \approx 1.054 \times 10^{-5} h^2 \; {\rm GeV/cm^3}$ is the critical energy density, and $h \approx 0.7$ is the rescaled Hubble parameter.

In the current scenario, the dark gauge boson $Z_D$ constitutes only a fraction of the total DM relic abundance. Thus, the differential photon flux from the decay and annihilation of $Z_D$ is now given by
\begin{align}
    \frac{d\Phi^{511}_\gamma}{d\Omega} =
    f_{Z_D}\frac{\Gamma_{Z_D\rightarrow e^-e^+}}{8\pi m_{Z_D}}R_\odot\rho_\odot J_{\rm decay}(\psi)
    +f_{Z_D}^2\frac{(\sigma v)_{Z_DZ_D\rightarrow e^-e^+}}{16\pi m_{Z_D}^2}
    R_\odot\rho_\odot^2 J_{\rm ann}(\psi)
    \,,
    \label{eqn:diffPhotonFlux-neq1over2}
\end{align}
where $f_{Z_D}$ is the fraction of the $Z_D$ relic in the total DM relic, {\it i.e.}, $f_{Z_D} \equiv \Omega_{Z_D}/\Omega_{\rm DM}$.

Before proceeding to the analysis of allowed parameter spaces, let us list the constraints that are relevant for our study. The constraints we consider are as follows:
\begin{itemize}
    \item{\bf DM relic density:} We consider the total DM relic density in the $3\sigma$ range put by the Planck collaboration \cite{Planck:2018vyg},
    \begin{align}
        0.1164 \leq \Omega_{\rm DM} h^2 \left( = \Omega_{Z_D} h^{2} + \Omega_{\psi_D} h^2 \right) \leq 0.1236\,.
    \end{align}

    \item{\bf Invisible Higgs decay and BSM Higgs bound:} We take into account the bound on the Higgs invisible branching ratio if it decays to dark sectors. Considering the DM in the MeV range, we have invisible decay of the SM Higgs boson. It is shown in Ref.~\cite{CMS:2022qva} that the allowed range of the SM Higgs branching ratio can maximally be ${\rm Br}\left(h_1 \rightarrow {\rm invisible}\right) \leq 0.18$. We consider this bound if a decay mode of the SM Higgs boson to the dark sector is kinematically open. Moreover, we also consider the bound on the Higgs mixing angle of $\sin\theta < 0.1$ in order to avoid the bounds associated with the Higgs signal strength and the beyond-the-SM (BSM) Higgs searches at the collider experiments looking for the SM fermions and vector bosons in the final states, produced from the BSM Higgs decay.

    \item{\bf Direct detection bound:} In the current study, the vector DM $Z_D$ can be detected by the direct detection experiments in contrary to the fermion DM which is dark to the visible sector.\footnote{
    There exist finite one-loop corrections to the $\psi_D N \rightarrow \psi_D N$ scattering, involving the dark $Z_D$ loop coupled to the dark Higgs boson, a mediator between the dark sector and the nucleons. We expect this contributions will have additional loop suppression factors compared to the tree-level contribution given in Eq.~\eqref{eqn:sidd-cs} and thus ignore it.
    }
    The spin-independent direct detection (SIDD) cross section can be expressed as
    \begin{align}
        \sigma_{\rm SI} = \frac{\mu^2 \sin^2(2\theta) g_D^2}{4 \pi v_H^2}
        \left( \frac{1}{m_{h_1}^2} - \frac{1}{m_{h_2}^2} \right)^2
        \left[ \frac{Z f^p_\alpha  + (A-Z) f^n_\alpha}{A} \right]^2
        \,,\label{eqn:sidd-cs}
    \end{align}
    where $\mu = m_{Z_D} m_N/(m_{Z_D} + m_N)$ with $m_N$ being the nucleon mass, and $Z$ and $A$ are the atomic and mass numbers, respectively. The proton and neutron form factors take the following values,
    \begin{align}
      f^p_\alpha &= m_N \left[
      \frac{7}{9} \left(f_p^u + f_p^d + f_p^s\right) + \frac{2}{9}
      \right]
      \,,\nonumber\\
      f^n_\alpha &= m_{N} \left[
      \frac{7}{9} \left(f_n^u + f_n^d + f_n^s\right) + \frac{2}{9}
      \right]
      \,,
    \end{align}
    where $f_p^u = f_n^d = 0.02$, $f_n^u = f_p^d = 0.026$, and $f_p^s = f_n^s = 0.043$ \cite{Junnarkar:2013ac}. 
    The relevant SIDD bounds for the mass range we consider in the current work come from XENONnT \cite{XENON:2023cxc}. Additionally, we also consider the limit coming from the neutrino floor.

    \item{\bf Indirect detection and 511 keV line:} For the model under consideration, the fermion DM only annihilates to the vector DM, and thus, there is no hope of detecting them at indirect detection experiments unless we take a relatively large value of the kinetic mixing angles $\epsilon$ and $\zeta$. On the other hand, the vector DM mainly annihilates to the dark Higgs boson, and annihilation to the SM fermions is subdominant. In general, the produced dark Higgs boson can decay to the SM fermions. However, we do not have active indirect searches going on for the mass range considered in the current study.
    On the contrary, we can have the Galactic 511 keV emission line signal from the decay and annihilation of the vector DM. The mass spectrum required by the consideration of the 511 keV emission line has the upper bound of 6 MeV (3 MeV) for the decay (annihilation) case. We mainly focus on the decay scenario and explore the allowed model parameters that give rise to the correct 511 keV photon flux, $(1/2)\times (0.96\pm0.07) \times 10^{-3} \; {\rm cm}^{-2}{\rm sec}^{-1}$ \cite{Siegert:2015knp}.

    \item{\bf $N_{\rm eff}$ bound:} In order to accompany the observed 511 keV emission line and to explain the correct DM relic abundance at the same time, we take our vector DM to be in the MeV scale. Therefore, the DM may freeze out during the Big Bang Nucleosynthesis (BBN) time and potentially alter the BBN prediction. While a more thorough and dedicated study is required to apply the BBN bound to the model under consideration, we may consider the general bound for the MeV-scale particle, which has been studied in, {\it e.g.}, Refs.~\cite{Chu:2023jyb, Sabti:2019mhn, EscuderoAbenza:2020cmq, Akita:2020szl, Bennett:2020zkv, Aloni:2023tff}; roughly, the mass less than 10 MeV is in conflict with $N_{\rm eff}$. It immediately seems to indicate that thermal MeV-scale DM cannot explain the 511 keV emission line. In the present work, however, we focus on the 511 keV emission line mostly produced from the DM decay, not from the annihilation of DM through the dark Higgs resonance, in which case the bound could be effective as the thermally-averaged DM cross section to $e^{+}e^{-}$ may be effective during the BBN time.
    Moreover, our DM increases the electron density indirectly when the produced dark Higgs boson decays to them during the BBN time. We expect that the increment in the electron density due to the DM freeze-out will be tiny, and we leave a dedicated study in this direction for the future. Nevertheless, we require, when appropriate, the dark Higgs boson to decay before the BBN.
\end{itemize}

We are now in a position to present the allowed parameter space that satisfies the above-mentioned bounds. We vary the model parameters in the following range:
\begin{gather}
  1 \leq m_{Z_D}\,[{\rm MeV}] \leq 100
  \,,\quad
  1 \leq m_{h_2}\,[{\rm MeV}] \leq 100
  \,,\nonumber\\
  10^{-5} \leq \theta \leq 0.1
  \,,\quad
  10^{-2} \leq g_D \leq \sqrt{4 \pi}
  \,,\quad
  10^{-25} \leq \epsilon, \zeta \leq 10^{-17}
  \,,\nonumber\\
  1 \leq m_{\psi_D}\,[{\rm GeV}] \leq 100
  \,,\quad
  10^{-2} \leq n_{\psi_D} g_{D} \leq \sqrt{4 \pi}
  \,.
\end{gather}
It is worth mentioning that we consider the vector DM in the MeV range and the fermion DM in the GeV range. If we were to take the fermion DM to be in the MeV range as well, we need to make sure that the vector DM mass is lower than the fermion DM mass, or the kinetic mixing angles are required to be sufficiently larger in order to make the annihilation of the fermion DM to the SM particles effective.

\begin{figure}[t!]
    \centering
    \includegraphics[scale=0.46]{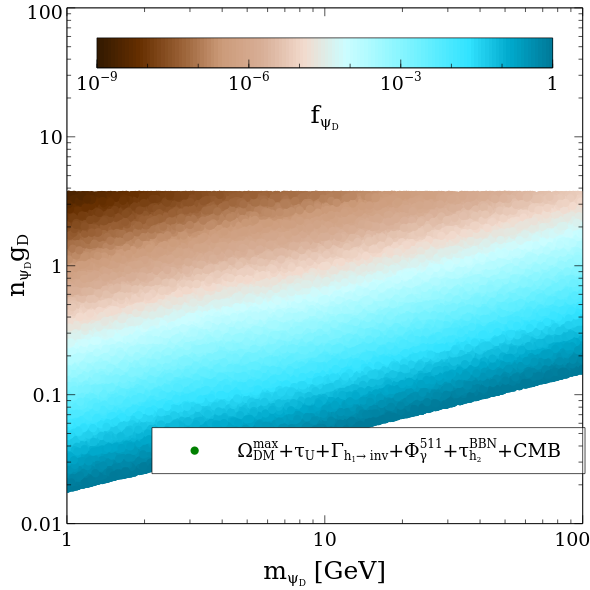}
    \;\;
    \includegraphics[scale=0.46]{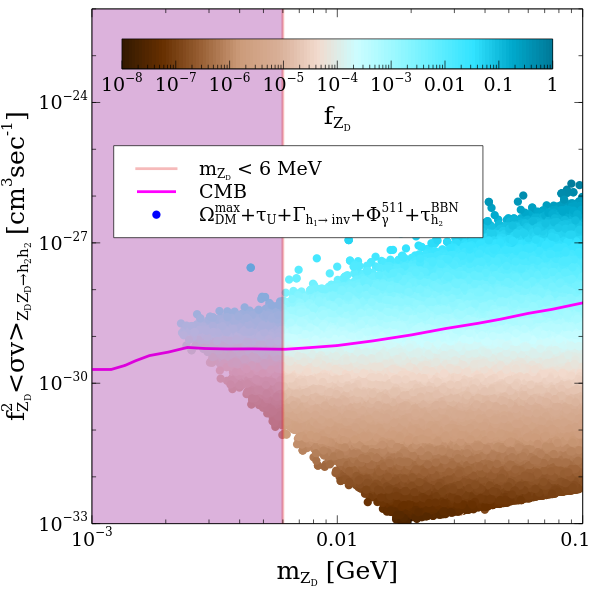}
    \caption{
      Allowed parameter space in the $m_{\psi_D}$--$n_{\psi_D} g_D$ plane (left) and $m_{Z_D}$--$f_{Z_D}^2 \langle \sigma v \rangle_{Z_D Z_D \rightarrow h_2 h_2}$ plane (right), satisfying the relic bound $\Omega^{\rm max}_{\rm DM} h^2 = 0.1236$, the condition that the lifetime of $Z_D$ is longer than the age of the Universe $\tau_{\rm U} \sim 10^{18}$ sec, the SM Higgs invisible decay bound $\Gamma_{h_1 \rightarrow {\rm inv}} < 0.18$, the observed Galactic 511 keV photon flux $\Phi_\gamma^{511}$, and the constraint that the dark Higgs decays before the BBN time $\tau^{\rm BBN}_{h_2} < 1$ sec. The left plot also takes into account the CMB bound \cite{Slatyer:2015jla} which is indicated as the magenta line in the right plot. The colours represent the fraction of the $\psi_D$ DM (left) and the fraction of the $Z_D$ DM (right). The shaded region shows the $m_{Z_D}<6$ MeV region that is required for the Galactic 511 keV emission line.}
    \label{fig:wimp-scatter-plot-1}
\end{figure}

In the left panel of Fig.~\ref{fig:wimp-scatter-plot-1}, we present the allowed parameter space in the $m_{\psi_D}$--$n_{\psi_D} g_D$ plane. The behaviour can easily be understood from Eqs.~\eqref{ab-psi-wd} and \eqref{eqn:YieldToday}. The general behaviour from the expressions indicate that if we increase $n_{\psi_D} g_{D}$, then the DM relic density will decrease, and if we increase the $m_{\psi_D}$, then the $\psi_D$ relic density will increase as the comoving number density effectively varies as $Y_{\psi_D} \propto m_{\psi}/(n_{\psi_D} g_D)^4$. This behaviour can be observed from the colour in the plot that represents the fraction of the $\psi_D$ relic density to the total DM density allowed by the Planck measurement. The important inference we can take from this observation is that $g_D$ may take any value to fix the relic density of the vector DM $Z_D$ and then we are free to change the $\psi_D$ charge $n_{\psi_D}$ accordingly in order to meet the correct value of total DM density. Therefore, in this subsection, we impose the condition that the $Z_D$ relic density does not exceed the upper bound on the DM relic density put by the Planck collaboration, $\Omega^{\rm max}_{\rm DM} h^2 = 0.1236$.
The right panel of Fig.~\ref{fig:wimp-scatter-plot-1} shows the allowed parameter space in the plane of $m_{Z_D}$--$f_{Z_D}^2 \langle \sigma v \rangle_{Z_D Z_D \rightarrow h_2 h_2}$ after imposing the relic bound $\Omega^{\rm max}_{\rm DM} h^2$, the condition that the lifetime of $Z_D$ is longer than the age of the Universe $\tau_{\rm U} \sim 10^{18}$ sec, the SM Higgs invisible decay bound $\Gamma_{h_{1} \rightarrow {\rm inv}} < 0.18$, the observed Galactic 511 keV photon flux $\Phi_\gamma^{511}$, and the constraint that the dark Higgs decays before the BBN time $\tau^{\rm BBN}_{h_2} < 1$ sec.
We see that the case where the $Z_D$ DM contributes 100\% of the total DM relic density is constrained by the CMB bound \cite{Slatyer:2015jla}. The sharp anti-correlation observed in the $m_{Z_D} \lesssim 10$ MeV is mainly due to the observed Galactic 511 keV photon flux. 

\begin{figure}[t!]
    \centering
    \includegraphics[scale=0.46]{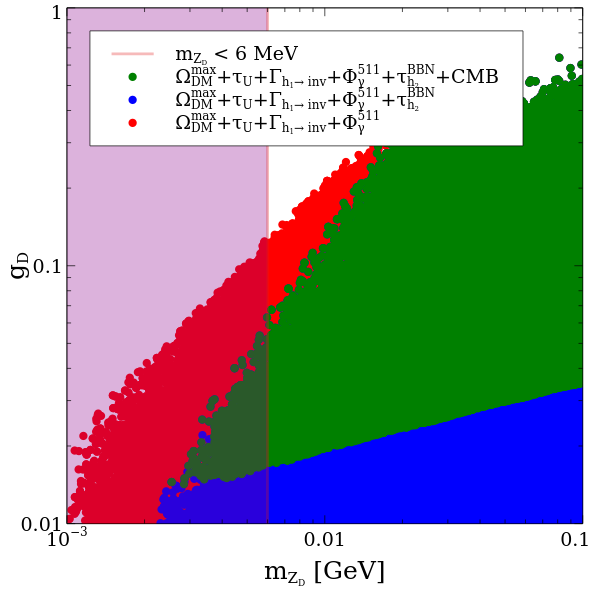}
    \;\;
    \includegraphics[scale=0.46]{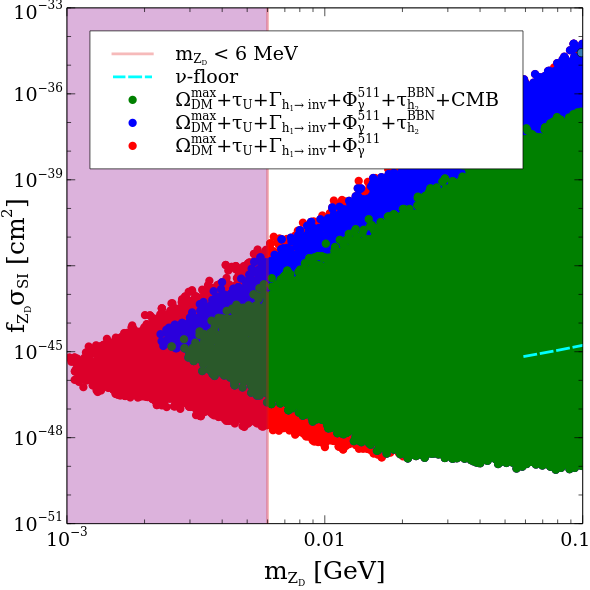}
    \caption{Allowed parameter space in the $m_{Z_D}$--$g_D$ plane (left) and $m_{Z_D}$--$f_{Z_D} \sigma_{\rm SI}$ plane (right). Different colours impose different sets of constraints. The red points are the results after imposing the relic bound $\Omega^{\rm max}_{\rm DM}h^2$, the $Z_D$ lifetime bound $\tau_{\rm U}$, the SM Higgs invisible decay bound $\Gamma_{h_1 \rightarrow {\rm inv}}$, and the observed Galactic 511 keV photon flux $\Phi_\gamma^{511}$. If we further impose the constraint that the dark Higgs decays before the BBN time, we obtain the blue points. The green points survive all the aforementioned bounds together with the CMB bound. The dashed cyan line indicates the neutrino floor, and the shaded region depicts the $m_{Z_D}<6$ MeV region.}
    \label{fig:wimp-scatter-plot-2}
\end{figure}

The left and right panels of Fig.~\ref{fig:wimp-scatter-plot-2} show the allowed parameter space in the $m_{Z_D}$--$g_D$ and $m_{Z_D}$--$f_{Z_D} \sigma_{\rm SI}$ planes, respectively. The colour variation in both plots implies the application of successive bounds; the red points are obtained after applying the relic bound $\Omega^{\rm max}_{\rm DM}$, the $Z_D$ lifetime bound $\tau_{\rm U}$, the SM Higgs invisible decay bound $\Gamma_{h_1 \rightarrow {\rm inv}}$, and the observed Galactic 511 keV photon flux $\Phi_\gamma^{511}$; the blue points are obtained if we further apply the constraint that the $h_2$ lifetime is smaller than the BBN time; and finally, the green points are obtained after applying the CMB bound together with all the aforementioned bounds.
The reduction of points after demanding that the $h_2$ lifetime is shorter than the BBN time of 1 second rules out $m_{h_2} \lesssim 2$ MeV and $10^{-5} < \sin\theta < 10^{-4}$, where the $h_2$ lifetime could become longer than 1 second and potentially cause problems with the BBN prediction.
On the other hand, imposing the CMB bound restricts higher $f_{Z_D}$ values, and it implies smaller values of $g_D$ as we discussed before.
Moreover, the SM Higgs invisible decay bound restricts the Higgs mixing angle to $\sin\theta \lesssim 10^{-3}$, and the CMB bound further restricts the mixing angle to $\sin\theta \lesssim 4\times 10^{-4}$.
We also note that the reduction of points after imposing the CMB bound in the right panel aligns well with the observation that the SIDD cross section \eqref{eqn:sidd-cs} is linearly proportional to the square of the mixing angle $\sin\theta$.

\begin{figure}[t!]
    \centering
    \includegraphics[scale=0.46]{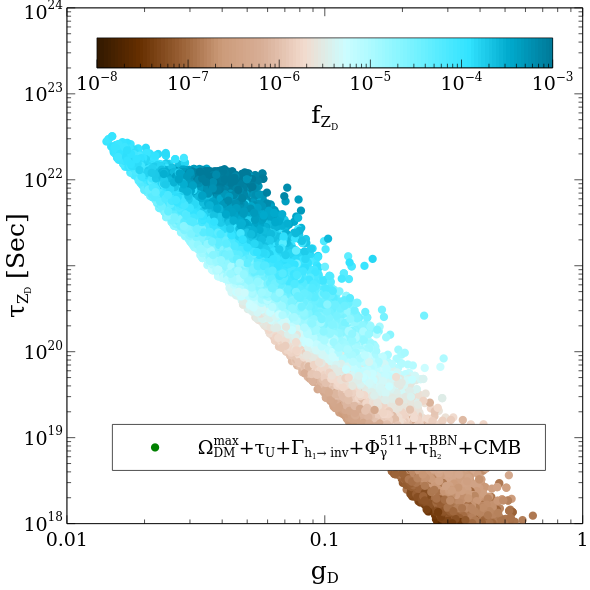}
    \;\;
    \includegraphics[scale=0.46]{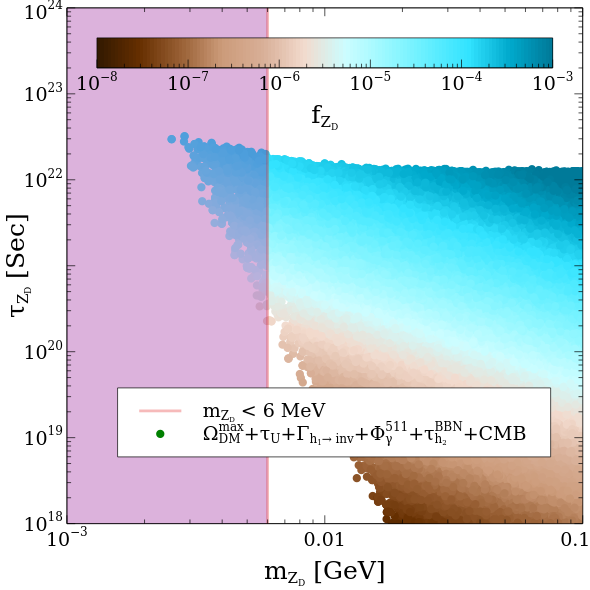}
    \caption{The $Z_D$ DM lifetime as a function of $g_D$ (left) and $m_{Z_D}$ (right). All the relevant bounds discussed earlier have been imposed. The colour depicts the fraction of the vector DM $f_{Z_D}$ to the total DM relic density, and the shaded region indicates the $m_{Z_D}<6$ MeV region.}
    \label{fig:wimp-scatter-plot-3}
\end{figure}

The left panel of Fig.~\ref{fig:wimp-scatter-plot-3} presents the lifetime of the vector DM $Z_D$ as a function of the dark gauge coupling $g_D$, after imposing all the relevant bounds discussed earlier. One may understand the result from Eq.~\eqref{eqn:diffPhotonFlux-neq1over2}; the flux is proportional to $f_{Z_D}/\tau_{Z_D}$ where the fraction of the $Z_D$ DM is inversely proportional to the square of the dark gauge coupling, $f_{Z_D} \propto g_D^{-2}$. Therefore, to keep the photon flux within the allowed range, the $Z_D$ lifetime $\tau_{Z_D}$ needs to be anti-correlated with $g_D$. We also observe that $\tau_{Z_D}$ becomes comparable to the age of the Universe $\tau_{\rm U}$ when $f_{Z_D}\sim 10^{-8}-10^{-7}$. The rest of the DM relic then comes from the fermion DM $\psi_D$, as we discussed before.
In the right panel, the $Z_D$ lifetime is shown as a function of $Z_D$ mass $m_{Z_D}$. As before, all the relevant bounds are applied. The shaded region indicates the $m_{Z_D} < 6$ MeV region that is relevant for the Galactic 511 keV emission line signal. The key lessons drawn from the plot are that $f_{Z_D}\sim 10^{-5}$ is required and that the $Z_D$ lifetime is many orders of magnitude longer than the age of the Universe.

Before we move on to the $n_{\psi_D} = 1/2$ case, it is also worth mentioning that we could study FIMP DM for the $n_{\psi_D} \neq 1/2$ case, where we can produce the vector DM from the decay of the Higgses and the fermionic DM from the annihilation of $Z_D$, namely $Z_{D}Z_{D} \rightarrow \psi_{D}\psi_{D}$. The production mode constrains the fermionic DM mass to be less than the vector DM mass. Thus, effectively, we do not have DM beyond the MeV scale if we wish to explain the 511 keV line. In addition, we also need to determine the distribution function of $Z_D$ for the production of $\psi_D$ from its annihilation. In some cases, a higher mass for $\psi_D$ is possible but we have to overcome two suppressions: one from non-thermal distribution and the other from the Boltzmann suppression due to the mass difference between the mother and daughter particles. 
This possibility may demand an extremely large value for the $U(1)_{D}$ charge of the fermionic DM, $n_{\psi_D}$.
This is different from the $n_{\psi_D} = 1/2$ case as we shall shortly see, where we may choose the value of the fermionic DM mass freely since it can be produced from the decay of the Higgses.
We also note that the difficulty in detections of FIMP-type DM can be relaxed by considering a low reheat temperature, which has been explored recently in, {\it e.g.}, Refs.~\cite{Cosme:2023xpa, Silva-Malpartida:2023yks}.
In the current work, we do not pursue the FIMP-type DM scenario for the $n_{\psi_D} \neq 1/2$ case.

\subsection{$n_{\psi_D} = 1/2$ Case}
\label{subsec:1over2Case}
Let us now move on to the $n_{\psi_D} = 1/2$ case. The Lagrangian is given by
\begin{align}
  \mathcal{L}_{1/2} = \mathcal{L}_{\rm M}
  + \bar{\psi}_D\left(i\gamma^\mu D_\mu - m_{\psi_D}\right)\psi_D
  -\frac{1}{\sqrt{2}}\left(
  y_{\psi_D}\phi_D^\dagger\overline{\psi_D^C} \psi_D
  + {\rm h.c.}
  \right)
  \,,
  \label{eqn:Lag1over2}
\end{align}
where $\mathcal{L}_{\rm M}$ is the Lagrangian for the minimal scenario, given in Eq.~\eqref{eqn:LagMinimal}, the covariant derivative is given by $D_\mu\psi_D = \partial_\mu\psi_D - (1/2) i g_D \hat{X}_\mu\psi_D$, and $y_{\psi_D}$ is the dark Yukawa coupling.
We decompose the dark fermion $\psi_D$ as
\begin{align}
  \psi_D = \frac{1}{\sqrt{2}}\left(\psi_D^R + i\psi_D^I\right)\,.
\end{align}
Then, upon spontaneous symmetry breaking of the dark $U(1)_D$, the masses of $\psi_D^{R,I}$ are given by
\begin{align}
    m_{\psi_D^{R,I}} = m_{\psi_D} \pm y_{\psi_D}v_D\,.
\end{align}
We assume a positive Yukawa coupling, so that $\psi_D^R$ is heavier than $\psi_D^I$. Thus, if the mass gap between them is reasonably small such that the lifetime of $\psi_D^R$ becomes longer than the age of the Universe, we may have a three-component DM scenario with the DM candidates being $Z_D$, $\psi_D^R$, and $\psi_D^I$. On the other hand, if the mass gap is large, a two-component DM scenario is obtained where $Z_D$ and $\psi_D^I$ become the DM candidates. These three- and two-component DM scenarios in this setup have been extensively studied in Ref.~\cite{Ahmed:2017dbb}.
In the present work, we focus on the two-component DM scenario and, unlike in Ref.~\cite{Ahmed:2017dbb}, consider the freeze-in production mechanism for the DM.
It is worth pointing out that, unlike the previous case of $n_{\psi_D} \neq 1/2$ where it is hard to detect the fermionic DM candidate at various DM searches through tree-level processes, the $n_{\psi_D} = 1/2$ case has much more fruitful detection prospects, which has been covered before in Ref.~\cite{Ahmed:2017dbb}. Detection of the fermionic DM for the $n_{\psi_D} \neq 1/2$ case may become possible via a one-loop process, but it will be suppressed by the loop effect and the Higgs mixing angle.

One may also have the possibility of explaining the 511 keV emission line with thermal DM. However, we do not expect any significant change in the conclusion compared to the previous study as the vector DM can only annihilate to $h_2$. The interesting aspects lie in the $\psi^{R,I}_{D}$ DM, which would have no effect on the 511 keV line apart from helping to satisfy the total DM density. Although there could be detection prospects for the fermionic DM component and some non-trivial effects in the fermionic DM production, there will be no change in the context of the 511 keV emission line. Therefore, in the current work, we refrain from exploring the WIMP scenario.

The Boltzmann equations for the dark sector particles, $Z_{D}$, $\psi_D^R$, and $\psi_D^I$ are given, in terms of the yields, by
\begin{align}
  \frac{d Y_{Z_{D}}}{d z} &=
  \frac{M_{\rm P} z \sqrt{g_{\rm eff}(z)}}{0.33 m_{h_1}^2 g_{*,s}(z)}
  \bigg[
  2\sum_{i=1,2} \langle \Gamma_{h_{i} \rightarrow Z_D Z_D} \rangle 
  \left( Y^{\rm eq}_{h_i} - Y_{Z_D}^2 \right) \theta(m_{h_i} - 2 m_{Z_D})
  \nonumber\\&
  +\langle \Gamma_{\psi_D^R \rightarrow \psi_D^I Z_D} \rangle_{\rm NTH} 
  \left( Y_{\psi_D^R} - Y_{\psi_D^I} Y_{Z_D} \right)
  \theta(m_{\psi_D^R} - m_{\psi_D^I} - m_{Z_D})
  \bigg]
  \,,\nonumber\\
  \frac{d Y_{\psi_D^{R(I)}}}{dz} &=
  \frac{M_{\rm P} z \sqrt{g_{\rm eff}(z)}}{0.33 m_{h_1}^2 g_{*,s}(z)}
  \bigg[
  2\sum_{i=1,2} \langle \Gamma_{h_i \rightarrow \psi_D^{R(I)} \psi_D^{R(I)}} \rangle 
  \left( Y^{\rm eq}_{h_i} - Y^2_{\psi_D^{R(I)}} \right)
  \theta(m_{h_i} - 2 m_{\psi_D^{R(I)}})
  \nonumber\\&
  -(+)\langle \Gamma_{\psi_D^R \rightarrow \psi_D^I Z_D} \rangle_{\rm NTH} 
  \left( Y_{\psi_D^R} - Y_{\psi_D^I} Y_{Z_D} \right)
  \theta(m_{\psi_D^R} - m_{\psi_D^I} - m_{Z_D})
  \bigg]
  \,,
  \label{eqn:be-fimp}
\end{align}
where $\langle\cdots\rangle$ and $\langle\cdots\rangle_{\rm NTH}$ are respectively the thermal and non-thermal averages, $\theta(x)$ is the Heaviside step function, $M_{\rm P}$ is the reduced Planck mass, $z = m_{h_1}/T$, and the effective degrees of freedom $g_{\rm eff}(z)$ is defined by \cite{Gondolo:1990dk,Edsjo:1997bg}
\begin{align}
  g_{\rm eff}(z) = \frac{g_{*,s}(z)}{\sqrt{g_*(z)}}
  \left( 1 - \frac{1}{3} \frac{d \ln g_{*,s}(z)}{d \ln z} \right)
  \,,
\end{align}
with $g_{*,s}(z)$ and $g_*(z)$ being the effective entropy and energy degrees of freedom, respectively.
The thermal and non-thermal averages of the decay width are given by \cite{Gondolo:1990dk,Konig:2016dzg,Biswas:2017ait}
\begin{align}
  \langle \Gamma_{h_i \rightarrow AA} \rangle =
  \Gamma_{h_i \rightarrow AA} 
  \frac{K_1(z)}{K_2(z)}
  \,,\quad
  \langle \Gamma_{\psi_D^R \rightarrow \psi_D^I Z_D} \rangle_{\rm NTH} =
  m_{\psi_D^R} \Gamma_{\psi_D^R \rightarrow \psi_D^I Z_D}
  \frac{\int \frac{\tilde{f}_{\psi_D^R} d^3p }{\sqrt{p^2 + m_{\psi_D^R}}}}{\int \tilde{f}_{\psi_D^R} d^3p}
  \,,
  \label{NTH-decay}
\end{align}
where $K_i(z)$ is the modified Bessel function of order $i$, and $\tilde{f}_{\psi_D^R}$ is the distribution function of $\psi_D^R$.
The expressions for the decay width $\Gamma_{h_{1,2} \rightarrow A A}$, where $A = \{Z_{D}, \psi_D^R, \psi_D^I\}$, and $\Gamma_{ \psi_D^R \rightarrow \psi_D^I Z_D}$ are given in Appendix~\ref{apdx:decaywidths}.
In the current study, we take the equilibrium distribution function for $\psi_D^R$ for simplicity, only to compute its non-thermal average of decay width as shown in Eq.~\eqref{NTH-decay}.
The production of $\psi_D^R$ happens through the decay of the Higgses which are in thermal equilibrium, and as such, the production of $\psi_D^R$ is unaffected by the choice of its distribution function.
In general, we can write down the non-thermal distribution function as $\tilde{f}_{\psi_D^R}(p) \propto \kappa(p) \tilde{f}^{\rm eq}{\psi_D^R}(p)$ \cite{Decant:2021mhj}, where $\kappa(p)$ is the momentum-dependent coefficient. 
The effect of the non-thermal distribution function may have different effects on the numerator and denominator \cite{Khan:2025yko}, and as a consequence, the decay lifetime can be affected for $z \lesssim 100$.
Proper treatment requires determining the non-thermal distribution function numerically, as shown in Refs.~\cite{Konig:2016dzg,Biswas:2017ait,Covi:2022hqb,Abdallah:2019svm}, where it is demonstrated that the shapes of the thermal and non-thermal distributions are almost identical, although they have different amplitudes. 
As we shall see below, however, in the present work, $\psi^R_D$ decays at around $z \sim 10^5$, where DM is non-relativistic and independent of any distribution. Thus, we may proceed without the need of the computation of the averaged decay width.

Once the solutions to the Boltzmann equations are given, we can determine the total DM density as
\begin{align}
  \Omega_{\rm DM} h^{2} =
  \Omega_{Z_D} h^2 + \Omega_{\psi_D^I} h^2 =
  \sum_{i = Z_D, \psi_D^I} 2.755 \times 10^8 Y_i \left(
  \frac{m_i}{\rm GeV}
  \right)
  \,.
\end{align}
It should be emphasized that the mass spectra are chosen in such a way that $Z_D$ and $\psi_D^I$ become suitable DM candidates. The main production modes of $\psi_D^I$ and $Z_D$ can be classified as the freeze-in and super-WIMP mechanisms. For the freeze-in production, we assume that the production of DM candidates is via the decay of the two Higgses $h_{1,2}$, depending on the kinematics. For the super-WIMP production, on the other hand, we consider the production from the decay of $\psi_D^R$.
Therefore, we may write
\begin{align}
  \Omega_{Z_D/\psi_D^I} h^2 =
  \Omega^{\rm FI}_{Z_D/\psi_D^I} h^{2} +
  \Omega^{\rm SW}_{Z_D/\psi_D^I} h^{2}
  \,,
\end{align}
where $\Omega^{\rm FI}_{Z_D/\psi_D^I}$ and $\Omega^{\rm SW}_{Z_D/\psi_D^I}$ represent the contributions from the freeze-in and super-WIMP mechanisms, respectively.
In general, for the process $A \rightarrow B C$, with $C$ being the DM, the freeze-in contribution $\Omega^{\rm FI}_C$ and the super-WIMP contribution $\Omega^{\rm SW}_C$ can be expressed as \cite{Hall:2009bx,Covi:2002vw}
\begin{align}
  \Omega^{\rm FI}_C h^2 =
  \frac{1.09 \times 10^{27}}{g_{*,s} \sqrt{g_*}} 
  \frac{m_C \Gamma_{A \rightarrow B C}}{m_A^2}
  \,,\qquad
  \Omega^{\rm SW}_C h^2 =
  \frac{m_C}{m_A} \Omega^{\rm FI}_A h^2
  \,. 
\end{align}
Taking $g_{*,s} = g_* = 100$, we analytically estimate the relic densities for $Z_D$ and $\psi_D$, up to the leading order in the gauge kinetic mixing angle, as
\begin{align}
    \Omega_{Z_D} h^2 &\approx
    1.08\times 10^{22} \times \sum_{i=1,2} c_i \frac{m_{Z_D}}{m_{h_i}v_D^2}
    \bigg[
    \left(\Delta m_{\psi_D}\right)^2
    \left(
    1 - \frac{4m_{\psi_D^R}^2}{m_{h_i}^2}
    \right)^{3/2}
    \theta\left( m_{h_i} - 2m_{\psi_D^R} \right)
    \nonumber\\&\quad
    +2m_{h_i}^2\sqrt{1 - \frac{4m_{Z_D}^2}{m_{h_i}^2}}
    \left(
    1 - \frac{4m_{Z_D}^2}{m_{h_i}^2} + \frac{12m_{Z_D}^4}{m_{h_i}^4}
    \right)
    \theta\left( m_{h_i} - 2m_{Z_D} \right)
    \bigg]\,,
    \label{omegazd-analytical}
\end{align}
and
\begin{align}
    \Omega_{\psi_D^I} h^2 &\approx
    1.08\times 10^{22} \times \sum_{i=1,2} c_i \frac{m_{\psi_D^I} (\Delta m_{\psi_D})^2}{m_{h_i}v_D^2}
    \bigg[
    \left(
    1 - \frac{4m_{\psi_D^R}^2}{m_{h_i}^2}
    \right)^{3/2}
    \theta\left( m_{h_i} - 2m_{\psi_D^R} \right)
    \nonumber\\&\quad
    +\left(
    1 - \frac{4m_{\psi_D^I}^2}{m_{h_i}^2}
    \right)^{3/2}
    \theta\left( m_{h_i} - 2m_{\psi_D^I} \right)
    \bigg]\,,
\end{align}
where $c_1 = \cos^2\theta$, $c_2 = \sin^2\theta$, and $\Delta m_{\psi_D} \equiv m_{\psi_D^R} - m_{\psi_D^I} = 2y_{\psi_D}v_D$.

\begin{figure}[t!]
    \centering
    \includegraphics[scale=0.46]{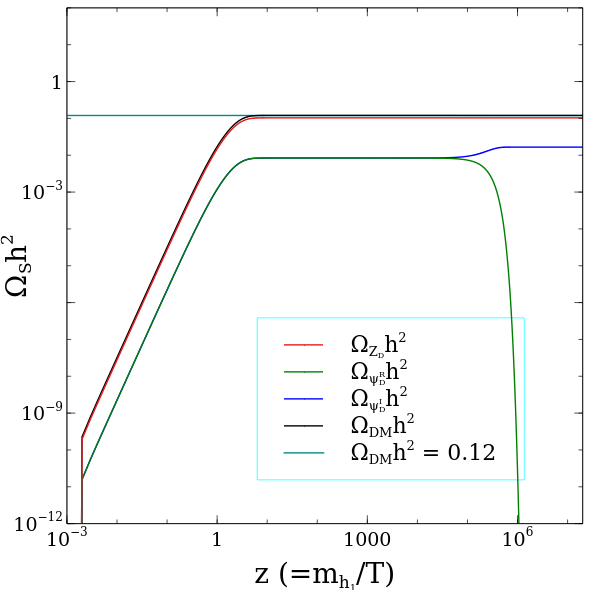}
    \;\;
    \includegraphics[scale=0.46]{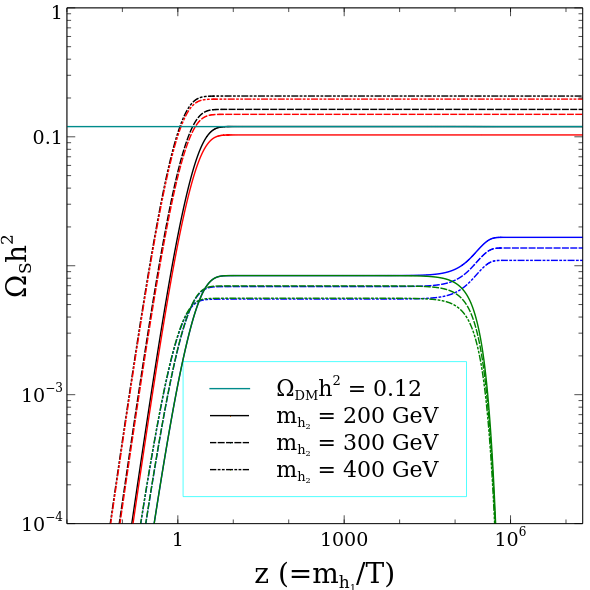}
    \caption{Evolution of the relic densities for the dark sector particles. The left panel shows the evolution of the $Z_D$ relic density (red), $\psi_D^R$ relic density (green), and $\psi_D^I$ relic density (blue) for the parameter values given in Eq.~\eqref{eqn:fit-params}. The black line indicates the total DM relic density, and the cyan line represents the Planck-observed value. The right panel shows the variation of the relic densities with regard to the change in the dark Higgs mass $m_{h_2}$. The other parameters are fixed to be those given in Eq.~\eqref{eqn:fit-params}.}
    \label{fig:line-plot-1}
\end{figure}

As an example, let us take the following parameter values:
\begin{gather}
  m_{Z_D} = 5 \, {\rm MeV}
  \,,\quad
  m_{\psi_D^R} = 50 \, {\rm GeV}
  \,,\quad
  m_{\psi_D^I} = 49 \, {\rm GeV}
  \,,\nonumber\\
  m_{h_2} = 200 \, {\rm GeV}
  \,,\quad
  \sin\theta = 0.1
  \,,\quad
  g_D = 1.08 \times 10^{-14}
  \,.\label{eqn:fit-params}
\end{gather}
This set of parameter values allows that all dark sector particles can be produced via the freeze-in mechanism from the $h_2$ as well as $h_1$ decays, if kinematically allowed.
The left panel of Fig.~\ref{fig:line-plot-1} shows the evolution of the relic densities for $Z_D$ (red), $\psi_D^R$ (green), and $\psi_D^I$ (blue). The black line indicates the total DM relic density, while the cyan line represents the Planck-observed value.
We clearly see from the plot the freeze-in production of the dark sector particles at around $z\sim 1$. Subsequently, the heavier dark fermion $\psi_D^R$ decays to $\psi_D^I$ and $Z_D$ DM at around $z\sim 10^5$, increasing their densities analogous to the super-WIMP mechanism.
A rise in the $\psi_D^I$ relic density is observed when $\psi_D^R$ decays, but no observable rise in $Z_D$ is seen due to the suppression factor arising from their mass ratio $m_{Z_D}/m_{\psi_D^R}$.
We estimate the $Z_D$ and $\psi_D^I$ relic densities to be $\Omega_{Z_D}h^2 = 0.1002$ and $\Omega_{\psi_D^I}h^2 = 0.0161$, which agrees with the results obtained after solving the Boltzmann equations \eqref{eqn:be-fimp}.

The right panel of Fig.~\ref{fig:line-plot-1} presents the evolution of relic densities for three different values of the dark Higgs mass $m_{h_2}$, namely 200 GeV (solid lines), 300 GeV (dashed lines), and 400 GeV (dot-dashed lines). The rest parameters are fixed as in Eq.~\eqref{eqn:fit-params}.
For vector DM production, the freeze-in contribution is proportional to the dark Higgs mass, which is evident from the red lines. On the other hand, for fermion DM production, shown by the blue lines, the production in both modes is inversely proportional to the dark Higgs mass, leading to a decrease in production with the increase of the dark Higgs mass.

\begin{figure}[t!]
    \centering
    \includegraphics[scale=0.46]{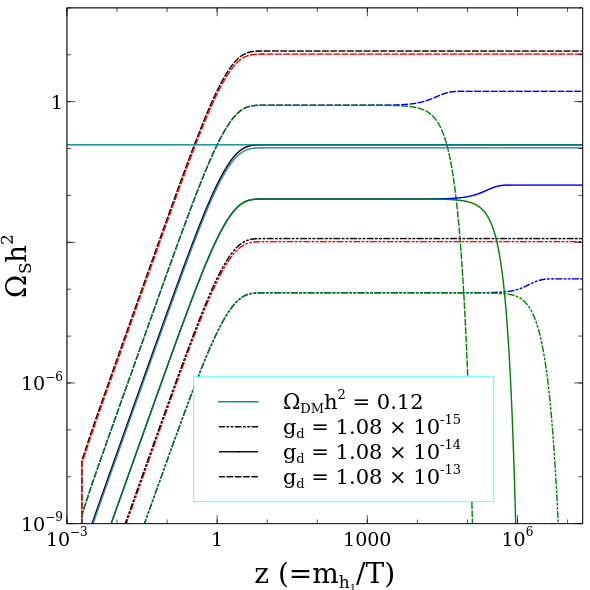}
    \;\;
    \includegraphics[scale=0.46]{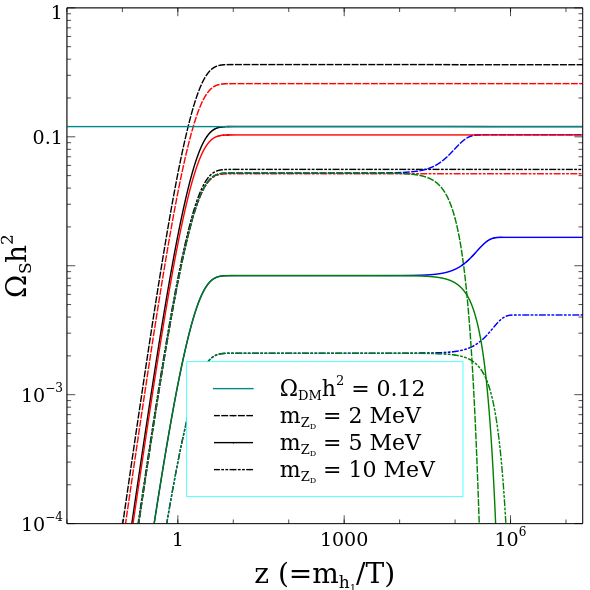}
    \caption{Variation of the relic densities with respect to the change in the gauge coupling $g_D$ (left) and the vector DM mass $m_{Z_D}$ (right). The other parameters are fixed to be those given in Eq.~\eqref{eqn:fit-params}.}
    \label{fig:line-plot-2}
\end{figure}

The variation of the evolution of the relic densities with respect to the change in the gauge coupling $g_D$ (the vector DM mass $m_{Z_D}$) is shown in the left (right) panel of Fig.~\ref{fig:line-plot-2}. From the analytical estimation, one may see that the fermion DM relic density and the vector DM relic density vary as $\Omega_{Z_D/\psi_D^I}h^2 \propto v_D^{-2} \propto g^2_D/m_{Z_D}^2$.
In the left panel, we see that as we increase the gauge coupling value, there is an increment in the production of both $\psi_D^I$ and $Z_D$, which is clear from the proportionality relation. On the other hand, as we increase the value of the vector DM mass, we see a decrement in the DM relic densities for both the DM components, which can also be seen from the analytical estimation.

\begin{figure}[t!]
    \centering
    \includegraphics[scale=0.46]{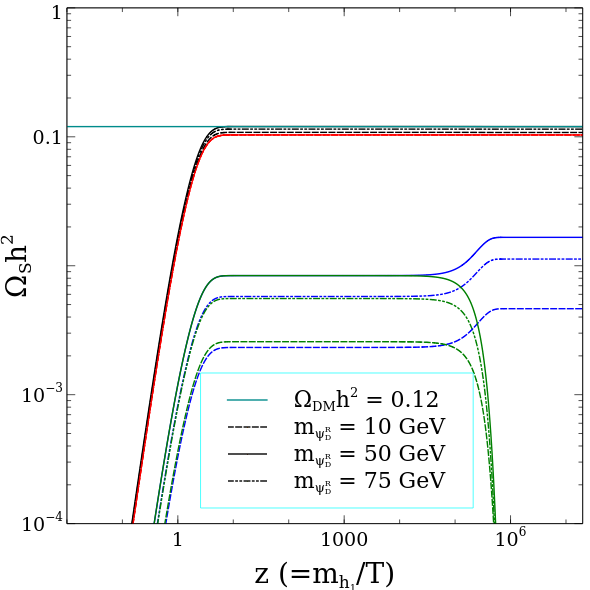}
    \;\;
    \includegraphics[scale=0.46]{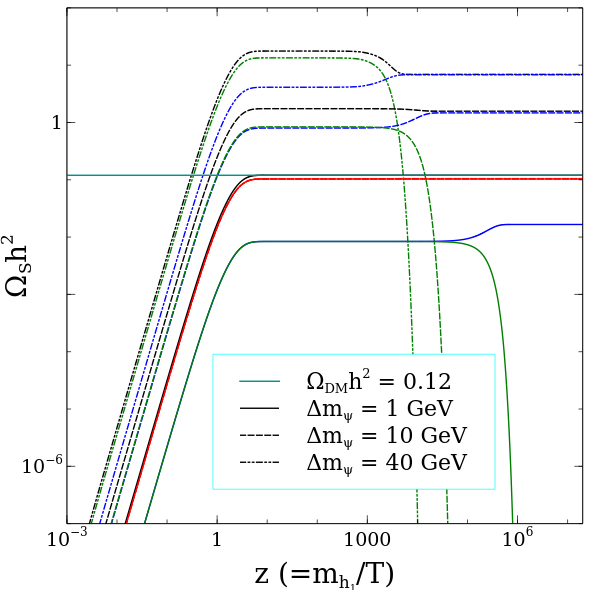}
    \caption{Variation of the relic densities with respect to the change in the $\psi_D^R$ mass (left) and $\Delta m_\psi \equiv m_{\psi_D^R} - m_{\psi_D^I}$ (right). The other parameters are fixed to be those given in Eq.~\eqref{eqn:fit-params}.}
    \label{fig:line-plot-3}
\end{figure}

In Fig.~\ref{fig:line-plot-3}, we present the effects of the change in the $\psi_D^R$ mass $m_{\psi_D^R}$ (left) and the mass difference $\Delta m_\psi \equiv m_{\psi_D^R} - m_{\psi_D^I}$ (left) on the DM relic density variations. From the left panel, we may learn that the change in the $\psi_D^R$ mass has a negligible impact on the vector DM production; this is because the freeze-in production is the dominant mode for the vector DM. On the other hand, for the fermion DM $\psi_D^I$, we see from the analytical estimation that the DM relic density varies as $\Omega_{\psi_D^I}h^2 \propto m_{\psi_D^I} ( 1 - 4 m_{\psi_D^I}^2/m_{h_2}^2 )^{3/2}$. As $m_{h_2} = 200$ GeV is chosen, we see from the figure that the relic density for $m_{\psi_D^I} = 50$ GeV is larger than the other two choices.
In the right panel, the DM relic density variations are shown for three different values of the mass difference $\Delta m_\psi$. The analytical estimation tells us that the fermion DM relic density varies as $(\Delta m_\psi)^2$ which is in good agreement with the behaviour shown in the plot. The vector DM also changes similarly due to the super-WIMP contribution but is suppressed due to its mass ratio $m_{Z_D}/m_{\psi_D^R}$. Therefore, we do not see any significant change in the vector DM relic density.

Having discussed the underlying physics of the freeze-in and super-WIMP DM production modes for our DM particles $Z_D$ and $\psi_D^I$, we now explore the allowed parameter space by performing a scan over the following ranges:
\begin{gather}
  1 \leq m_{Z_D}\,[{\rm MeV}] \leq 100
  \,,\quad
  1 \leq m_{\psi_D^I}\,[{\rm GeV}] \leq 100
  \,,\quad
  m_{\psi_D^I} \leq m_{\psi_D^R} \leq m_{\psi_D^I} \times 10^3
  \,,\nonumber \\
  125 \leq m_{h_2}\,[{\rm GeV}] \leq 10^4
  \,,\quad
  10^{-5} \leq \theta \leq 10^{-1}
  \,,\nonumber \\
  10^{-18} \leq g_D \leq 10^{-12}
  \,,\quad
  10^{-25} \leq \epsilon,\zeta \leq 10^{-17}
  \,.
\end{gather}
We impose the DM relic density bound at $3\sigma$, the perturbative and stability constraints on the quartic couplings ($\lambda_{hH} > 0$ is considered here), and the condition that the $Z_D$ lifetime is longer than the age of the Universe $\tau_{\rm U}\sim 10^{18}$ sec:
\begin{align}
  0.1164 \leq \Omega_{Z_D} h^2 + \Omega_{\psi_D^I} h^2 \leq 0.1236
  \,,\quad
  0 < \lambda_{h,H,hH} < 4 \pi
  \,,\quad
  \tau_{Z_D} > \tau_{\rm U}
  \,.
\end{align}
On top of these constraints, we select parameter sets that give rise to the observed 511 keV photon flux. We note that, at this point, only the value of the photon flux is imposed, and the condition that $m_{Z_D} < 6$ MeV, which is required to account for the Galactic 511 keV emission line signal, is not demanded. The corresponding $m_{Z_D} < 6$ MeV region shall be indicated later.

\begin{figure}[t!]
    \centering
    \includegraphics[scale=0.46]{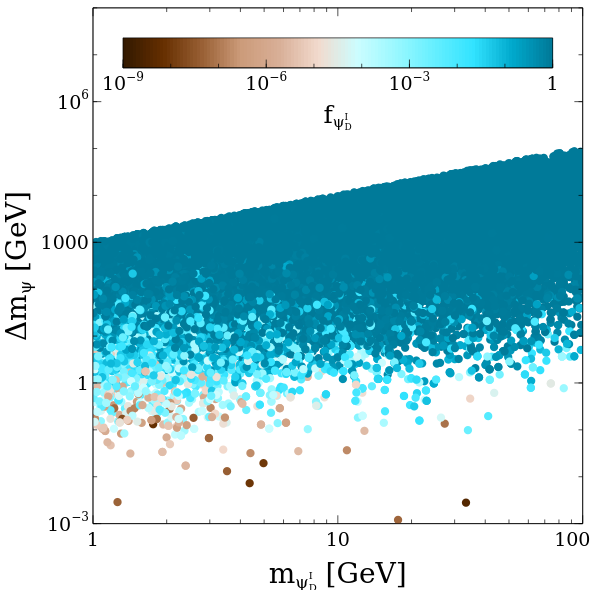}
    \;\;
    \includegraphics[scale=0.46]{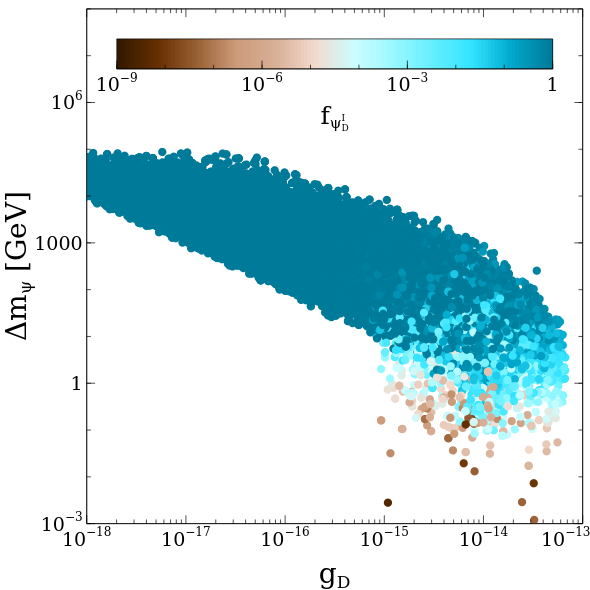}
    \caption{Allowed parameter space in the $m_{\psi_D^I}$--$\Delta m_\psi$ (left) and $g_D$--$\Delta m_\psi$ (right) planes, satisfying the DM relic density bound, the perturbative and stability constraints, the condition that the $Z_D$ lifetime is longer than the age of the Universe, and the observed 511 keV photon flux. We note that only the value of the photon flux is imposed, without the requirement of $m_{Z_D} < 6$ MeV, which is needed to account for the Galactic 511 keV emission line signal.}
    \label{fig:scatter-plot-1}
\end{figure}
  
In Fig.~\ref{fig:scatter-plot-1}, we present the allowed parameter space in the $m_{\psi_D^I}$--$\Delta m_\psi$ (left) and $g_D$--$\Delta m_\psi$ (right) planes; all the aforementioned constraints are satisfied. In both the left and right panels, the colour represents the fraction of the fermion DM $\psi_D^I$, {\it i.e.}, $f_{\psi_D^I}$.
The left panel shows that $f_{\psi_D^I}$ increases as the value of $\Delta m_\psi$ increases, which can be understood from our analytical estimation that the relic density of $\psi_D^I$ is proportional to the square of the mass difference. The sharp upper cut for $\Delta m_\psi$ represents the maximum value allowed in our parameter scan. We also see that, for the $\Delta m_\psi < 10$ GeV region, DM mostly consists of the vector DM.
The right panel indicates that, for $g_D < 10^{-15}$, DM is mostly dominated by the fermion DM $\psi_D^I$, and this region corresponds to $\Delta m_\psi > 10$ GeV. On the other hand, for $10^{-15} \leq g_D \leq 10^{-13}$ and $\Delta m_\psi < 10$ GeV, the fermion DM $\psi_D^I$ is subdominant, and most of the contribution comes from vector DM $Z_D$.

\begin{figure}[t!]
    \centering
    \includegraphics[scale=0.46]{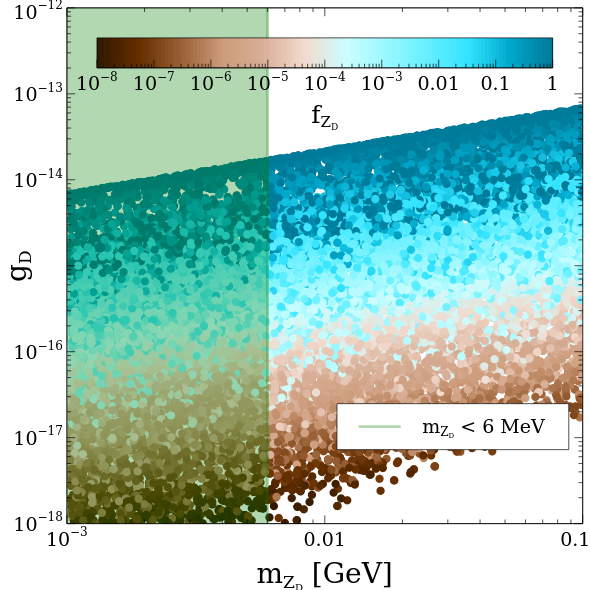}
    \;\;
    \includegraphics[scale=0.46]{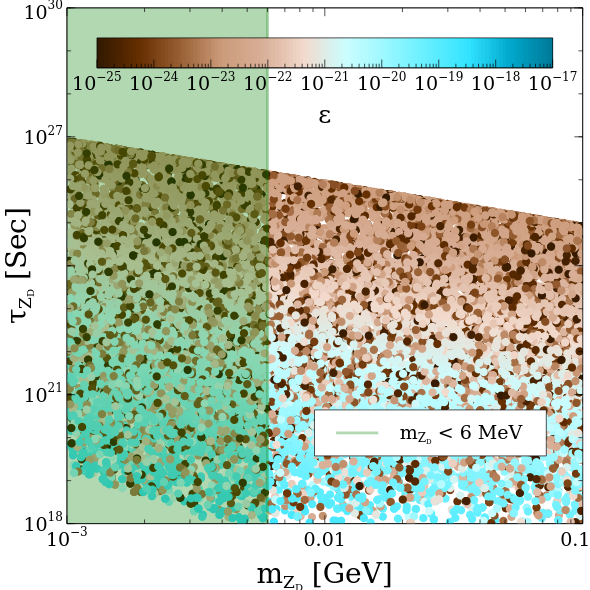}
    \caption{Allowed parameter space in the $m_{Z_D}$--$g_D$ (left) and $m_{Z_D}$--$\tau_{Z_D}$ (right) planes, satisfying the aforementioned constraints. The green-shaded region indicates the $m_{Z_D} < 6$ MeV region which is suitable for explaining the observed Galactic 511 keV line signal. In the left panel, the colour variation shows the fraction of vector DM, whereas in the right panel, the colour represents the value of the gauge kinetic mixing angle.}
    \label{fig:scatter-plot-2}
\end{figure}

The left panel of Fig.~\ref{fig:scatter-plot-2} presents the allowed parameter space in the $m_{Z_D}$--$g_D$ plane, where the colour variation shows the fraction of the vector DM compared to the total DM relic density. We note that all the points shown in the plot satisfy the aforementioned constraints.
The vector DM relic is inversely proportional to the square of the dark Higgs VEV, $\Omega_{Z_{D}}h^{2} \propto 1/v^2_{D} \propto g^2_{D}/m^2_{Z_D}$, as we learnt from the analytical estimation in Eq.~\eqref{omegazd-analytical}.
Therefore, the vector DM relic density fraction $f_{Z_D}$ directly proportional to the square of the gauge coupling $g_D$ when a fixed dark gauge boson mass $m_{Z_{D}}$ is considered. This tendency is clearly observed in the colour variation.
The upper sharp line in the allowed range is due to the upper limit of the DM relic density. We note that any fraction of the vector DM relic $f_{Z_D}$ is possible for any value of the vector DM mass $m_{Z_D}$.
The right panel of Fig.~\ref{fig:scatter-plot-2} shows the $Z_D$ lifetime as a function of the vector DM mass $m_{Z_D}$. The colour depicts the value of the gauge kinetic mixing angle $\epsilon$. We observe that the upper line descends as $m_{Z_D}$ increases, which is opposite to the trend seen in the left panel; this is because $\tau_{Z_D}$ is inversely proportional to the square of the gauge coupling. We also observe that a large range of $\epsilon$ values is allowed near the age of the Universe, $\tau_{Z_D} \sim 10^{18}$ sec, as smaller values of $\epsilon$ can be compensated by higher values of the other mixing angle $\zeta$.
In both the left and right panels, the green-shaded region corresponds to the region that can provide the correct energy spectrum of the Galactic 511 keV signal.
Although the entire range of the vector DM fraction seems to be allowed, we shall shortly show that very small $f_{Z_D}$ values give rise to the $Z_D$ lifetime to be of the same order as the age of the Universe, {\it i.e.}, $\tau_{Z_D}\sim \tau_{\rm U}$, while large $f_{Z_D}$ values result in $\tau_{Z_D}\gg\tau_{\rm U}$.

\begin{figure}[t!]
    \centering
    \includegraphics[scale=0.46]{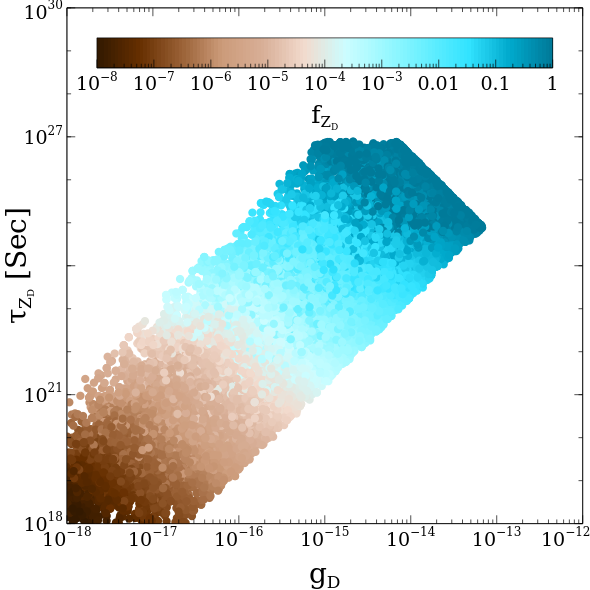}
    \;\;
    \includegraphics[scale=0.46]{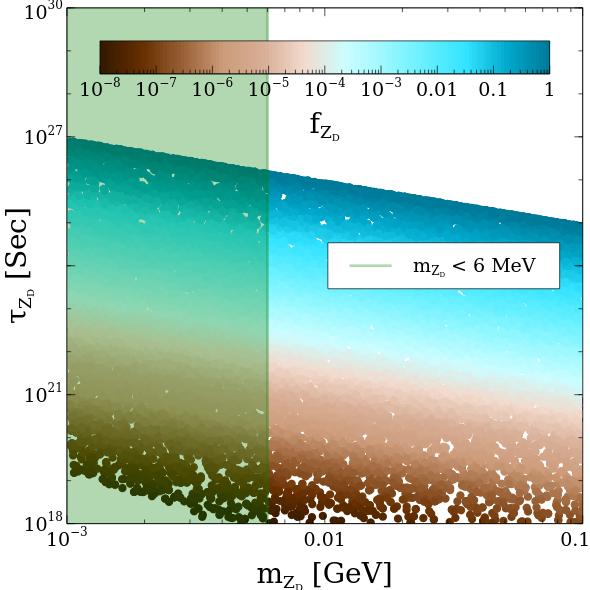}
    \caption{The lifetime of the vector DM $Z_D$ in terms of the gauge coupling $g_D$ (left) and the vector DM mass $m_{Z_D}$ (right). As before, all the aforementioned constraints are satisfied. In both the left and right panels, the colour represents the fraction of the vector DM relic $f_{Z_D}$. The green-shaded region in the right panel indicates the $m_{Z_D} < 6$ MeV region.}
    \label{fig:scatter-plot-3}
\end{figure}

The left and right panels of Fig.~\ref{fig:scatter-plot-3} show the $Z_D$ lifetime as a function of the gauge coupling $g_D$ and the vector DM mass $m_{Z_D}$, respectively. The colour represents the fraction of the vector DM relic $f_{Z_D}$ in both the left and right panels.
Interestingly, compared to the left panel of Fig.~\ref{fig:wimp-scatter-plot-3}, the left panel of Fig.~\ref{fig:scatter-plot-3} features a different behaviour; this is because, in the current case, the relic density is linearly proportional to the square of the gauge coupling, as we may see from our analytical estimation.
It is clear that the $Z_D$ lifetime is proportional to the fraction of the vector DM relic $f_{Z_D}$. In other words, for $\tau_{Z_D} \sim \tau_{\rm U}$, $f_{Z_D}$ is required to be tiny, while large values of $f_{Z_D}$ result in $\tau_{Z_D} \gg \tau_{\rm U}$.
It is also interesting to note that, unlike the $n_{\psi_D} \neq 1/2$ case, $f_{Z_D} = 1$ becomes a possibility in the $m_{Z_D} < 6$ MeV region. The $Z_D$ lifetime needs to be significantly longer than the age of the Universe in this case. Therefore, if we consider freeze-in production, we do not need the fermionic DM component to explain the 511 keV line and account for the total DM density. 
However, the presence of the fermionic DM provides a much wider region of the parameter space that can explain the 511 keV line as well as the total DM density observed from the Planck satellite. Only a narrow parameter region would be allowed if the fermionic DM was not present.
We note that this situation is different from the WIMP DM case with $n_{\psi_D} \neq 1/2$ where the fermionic DM is very much needed as we do not have 100\% vector DM after satisfying all the bounds and the 511 keV signal. In the case of $n_{\psi_D} = 1/2$, the presence of the fermionic DM is not essential but required to obtain a wider allowed parameter space.

\begin{figure}[t!]
    \centering
    \includegraphics[scale=0.46]{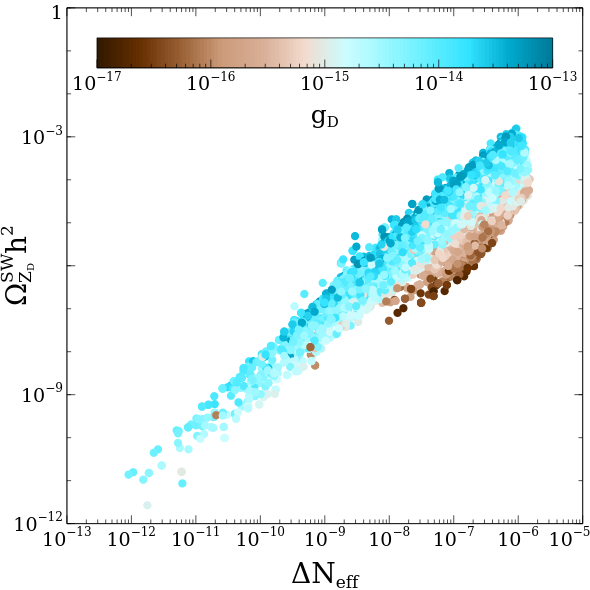}
    \;\;
    \includegraphics[scale=0.46]{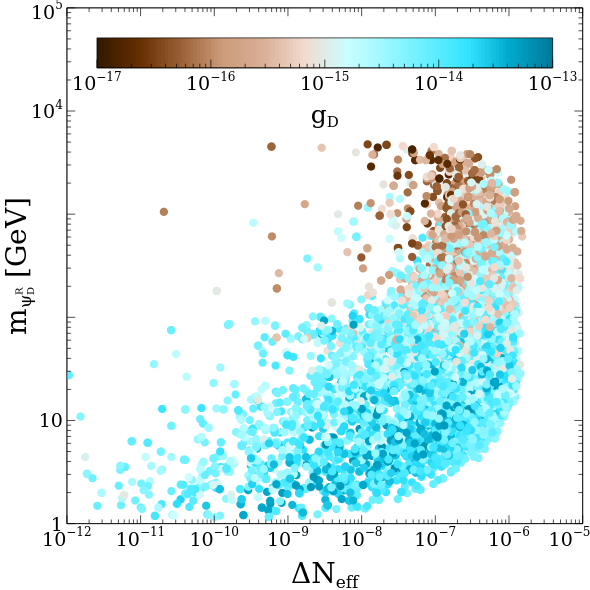}
    \caption{The left (right) panel shows the correlation between $\Delta N_{\rm eff}$, the effective relativistic degrees of freedom, and $\Omega_{Z_D}^{\rm SW}h^2$, the relic density of $Z_D$ produced via the late decay of $\psi_D^R$ ($m_{\psi_D^R}$, the mass of $\psi_D^R$). As before, all the aforementioned constraints are satisfied, and we have also demanded that all the points satisfy the correct value of the photon flux but with a varied energy spectrum. In both left and right panels, the colour represents the value of the dark gauge coupling $g_D$.}
    \label{fig:scatter-plot-4}
\end{figure}

Before we conclude, let us discuss the possible contribution of the DM, produced by the late decay of the heavier dark fermion $\psi_D^R$, to the relativistic degrees of freedom. Closely following Ref.~\cite{Decant:2021mhj}, the decay width of $\psi_D^R$ to the DM particles $Z_D$ and $\psi_D^I$ can be related to the effective relativistic degrees of freedom $\Delta N_{\rm eff}$ as
\begin{align}
  \Delta N_{\rm eff} =
  1.2 \times 10^{-2} \,
  \delta_{Z_D} \,
  \left( \frac{m_{\psi_D^R}}{100\,{\rm GeV}} \right)
  \left( \frac{10^{-22}\,{\rm GeV}}{\Gamma_{\psi_D^R \rightarrow \psi_D^I Z_D}} \right)^{1/2}
  \left( \frac{\Omega_{Z_D} h^2}{0.12} \right)
  \left( \frac{1\,{\rm MeV}}{m_{Z_D}} \right)
  \,,
  \label{eqn:neff}
\end{align}
and similarly for $\psi_D^I$, where $\delta_{Z_D} = (m^2_{\psi_D^R} - m_{\psi_D^I}^2 )/m_{\psi_D^R}^2$.
The left panel of Fig.~\ref{fig:scatter-plot-4} presents the correlation between the relic density of $Z_D$ produced via the super-WIMP mechanism $\Omega^{\rm SW}_{Z_D}h^2$ and the effective relativistic degrees of freedom $\Delta N_{\rm eff}$ with the colour indicating the value of the gauge coupling $g_D$. Here, we have imposed the condition that $m_{h_2} > 2 m_{\psi_D^R}$ so that $\psi_D^R$ could always be produced through the freeze-in mechanism and then later decay to the DM candidates. Due to the negligible contribution of the late decay of $\psi_D^R$ to the $Z_D$ relic density, we observe that $\Delta N_{\rm eff}$ remains very small. We further note that, for a fixed value of $\Omega^{\rm SW}_{Z_D}h^2$, $\Delta N_{\rm eff}$ increases as the gauge coupling $g_D$ decreases. This behaviour aligns well with our understanding from Eq.~\eqref{eqn:neff} that larger $g_D$ would cause more effective late decay of $\psi_D^R$.
The right panel shows the correlation between the $\psi_D^R$ mass and $\Delta N_{\rm eff}$. The colour again denotes the value of $g_D$. From the plot, we see that higher values of $m_{\psi_D^R}$ correspond to lower values of $g_D$; this behaviour mainly comes from the DM relic density bound. Similarly to the left panel, we also see that small $g_D$ values tend to lead to larger values of $\Delta N_{\rm eff}$.
The situation remains very similar for the fermion DM $\psi_D^I$; in this case, as one may see from Eq.~\eqref{eqn:neff}, while we can have a boost from the mass difference factor $\delta_{\psi_D^I}$, we have a suppression factor of $1/m_{\psi_D^I}$ as $m_{\psi_D^I}$ is in the GeV range.
It is worth mentioning that the detection of such a small variation in the effective relativistic degrees of freedom would be challenging in future experiments.

\section{Conclusion}
\label{sec:conc}
In this paper, we have revisited the line signal of the Galactic 511 keV $\gamma$-ray emission in the framework of a local, dark $U(1)_D$ extension of the Standard Model. The dark $U(1)_D$ naturally accommodates a gauge boson which may decay and/or pair-annihilate to the electron-positron pairs due to the presence of the gauge kinetic mixing and the dark Higgs boson. The produced positrons form positronium that eventually annihilates to two photons, sourcing the observed Galactic 511 keV $\gamma$-ray emission.
The model also naturally provides dark matter candidates and may account for the correct dark matter relic density. In the case of the minimal scenario where only the dark gauge boson and the dark Higgs boson are introduced, the dark gauge boson plays the role of dark matter. We have shown how the vector dark matter may produce the electron-positron pairs, computed the photon flux, and compared with the observed value.

Upon introducing a dark fermion to the minimal setup, a multi-component dark matter scenario could be realised. In this case, both the dark fermion and the dark gauge boson become good dark matter candidate. The minimal scenario could then be viewed as a limiting case of the multi-component dark matter scenario. Depending on the $U(1)_D$ charge of the dark fermion, one may categorise the model into two; one where the charge is not equal to 1/2 and the other where the charge is 1/2, in which case a Yukawa interaction term can be admitted. We have considered both cases in the present work. For the former, we have explored the WIMP-type dark matter scenario with the standard freeze-out mechanism. For the latter, the FIMP-type dark matter scenario has been discussed. In this case, both the vector and fermion dark matter particles could be produced via the freeze-in and super-WIMP mechanisms. The minimal scenario could then correspond to the case where the vector dark matter constitutes the 100\% of the total dark matter.

Taking into account various theoretical and experimental constraints, we have explored the parameter space that successfully accounts not only for the Galactic 511 keV $\gamma$-ray emission but also for the correct dark matter relic density. For the WIMP scenario, we have demonstrated that the case where the vector dark matter becomes the 100\% of the total dark matter, which thus corresponds to the minimal scenario, is incapable of explaining both the correct relic density and the 511 keV signal mainly due to the stringent CMB bound. The fraction of the vector dark matter has been found to be smaller than $10^{-3}$. For the FIMP scenario, on the other hand, the fraction of the vector dark matter could become unity, while satisfying all the relevant constraints. However, the lifetime of the vector dark matter turned out to be many orders of magnitude longer than the age of the Universe. Similarly to the WIMP case, the vector dark matter could constitute a small fraction of the total dark matter, in which case the lifetime may become comparable to the age of the Universe. We have also discussed possible impacts of the dark matter on the effective relativistic degrees of freedom in terms of $N_{\rm eff}$ and found that the contribution is small.

While the model we have considered in the present work is successful at explaining the correct dark matter relic density and the observed line signal of the Galactic 511 keV $\gamma$-ray emission, it may be challenging to test the model with other means. For the WIMP case, the freeze-out of the light vector dark matter may occur during the BBN time in which case the BBN prediction can potentially be altered. A dedicated study is necessary with a multi-component dark matter setup. For the FIMP case, the lifetime of the vector dark matter may become comparable to the age of the Universe. Thus, it may leave imprints on large scale structures, and future Galaxy-scale surveys can further provide a potential probe. Finally, we have taken into account only the magnitude of the observed photon flux without considering the angular distribution. It has been argued that the morphology of the observed photon flux favours annihilating dark matter scenarios over decaying scenarios \cite{Ascasibar:2005rw,Prantzos:2010wi,Vincent:2012an}, although it does not completely rule out the decaying dark matter due to the uncertainties in the astrophysical dynamics. Therefore, a detailed study on the morphology for the model considered in the present work would be valuable.
We plan to come back to these questions in the future.

\acknowledgments
The authors contributed equally to this work and are listed in alphabetical order.
This work was supported in part by Basic Science Research Program through the National Research Foundation of Korea funded by the Ministry of Education, Science and Technology (NRF-2022R1A2C2003567 and RS-2024-00341419) (JKK) and by KIAS Individual Grants under Grant No. PG021403 at Korea Institute for Advanced Study (PK).
This research was also supported in part by Brain Pool program funded by the Ministry of Science and ICT through the National Research Foundation of Korea (RS-2024-00407977) (SK).

\appendix
\section{Expressions for Decay Widths}
\label{apdx:decaywidths}
We summarise expressions for the decay widths relevant for the current study.
The decay width for the process $\psi_D^R \rightarrow \psi_D^I Z_D$ is given by
\begin{align}
    \Gamma_{\psi_D^R \rightarrow \psi_D^I Z_D} &=
    \frac{g_D^2 \cos^2\zeta}{32\pi\cos^2\epsilon} m_{\psi_D^R}\left(\frac{m_{Z_D}}{m_{\psi_D^R}}\right)^4
    \sqrt{\left[
    \left(\frac{\Delta m_{\psi_D}}{m_{Z_D}}\right)^2-1
    \right]\left[
    \left(\frac{\Delta m_{\psi_D}+2m_{\psi_D^I}}{m_{Z_D}}\right)^2-1
    \right]}
    \nonumber\\ &\qquad\times
    \left[
    \left(\frac{\Delta m_{\psi_D}}{m_{Z_D}}\right)^2-1
    \right]\left[
    1+\frac{(\Delta m_{\psi_D}+2m_{\psi_D^I})^2}{2m_{Z_D}^2}
    \right]
    \,.
\end{align}
where $\Delta m_{\psi_D} \equiv m_{\psi_D^R}-m_{\psi_D^I} = 2y_{\psi_D}v_D$. \\
The decay width for the process $h_2 \rightarrow Z_D Z_D$ is given by
\begin{align}
    \Gamma_{h_2 \rightarrow Z_D Z_D} &=
    \frac{\cos^2\theta \cos^4\zeta g_D^4 v_D^2 m_{h_2}^3}{32\pi \cos^4\epsilon m_{Z_D}^4}
    \sqrt{1-\frac{4m_{Z_D}^2}{m_{h_2}^2}}
    \left(
    1-\frac{4m_{Z_D}^2}{m_{h_2}^2}+\frac{12m_{Z_D}^4}{m_{h_2}^4}
    \right)
    \nonumber\\ &\quad \times 
    \left[
    1
    -\frac{e^2\tan\theta v_H}{4\cos^2\theta_w \sin^2\theta_w g_D^2 v_D}\left(\sin\epsilon\sin\theta_w - \cos\epsilon\tan\zeta\right)^2
    \right]^2\,,
\end{align}
while for the process $h_1 \rightarrow Z_D Z_D$, we have
\begin{align}
    \Gamma_{h_1 \rightarrow Z_D Z_D} &=
    \frac{\sin^2\theta \cos^4\zeta g_D^4 v_D^2 m_{h_1}^3}{32\pi \cos^4\epsilon m_{Z_D}^4}
    \sqrt{1-\frac{4m_{Z_D}^2}{m_{h_1}^2}}
    \left(
    1-\frac{4m_{Z_D}^2}{m_{h_1}^2}+\frac{12m_{Z_D}^4}{m_{h_1}^4}
    \right)
    \nonumber\\ &\quad \times 
    \left[
    1
    +\frac{e^2 v_H}{4\cos^2\theta_w \sin^2\theta_w \tan\theta g_D^2 v_D}\left(\sin\epsilon\sin\theta_w - \cos\epsilon\tan\zeta\right)^2
    \right]^2\,.
\end{align}
Finally, the decay widths for the processes $h_{1,2} \rightarrow \psi_D^{R/I} \psi_D^{R/I}$ are given by
\begin{align}
  \Gamma_{h_2 \rightarrow \psi_D^{I/R}\psi_D^{I/R}} &=
  \frac{\cos^2\theta m_{h_2} y_{\psi_D}^2}{16\pi}\left(
  1-\frac{4m_{\psi_D^{I/R}}^2}{m_{h_2}^2}
  \right)^{3/2}
  \,,\nonumber\\
  \Gamma_{h_1 \rightarrow \psi_D^{I/R}\psi_D^{I/R}} &=
  \frac{\sin^2\theta m_{h_1} y_{\psi_D}^2}{16\pi}\left(
  1-\frac{4m_{\psi_D^{I/R}}^2}{m_{h_1}^2}
  \right)^{3/2}
  \,.
\end{align}


\bibliographystyle{JHEP}
\bibliography{main}

\providecommand{\href}[2]{#2}\begingroup\raggedright\begin{thebibliography}{10}

\bibitem{Planck:2018vyg}
{\scshape Planck} collaboration, \emph{{Planck 2018 results. VI. Cosmological
  parameters}},
  \href{https://doi.org/10.1051/0004-6361/201833910}{\emph{Astron. Astrophys.}
  {\bfseries 641} (2020) A6}
  [\href{https://arxiv.org/abs/1807.06209}{{\ttfamily 1807.06209}}].

\bibitem{Profumo:2019ujg}
S.~Profumo, L.~Giani and O.F.~Piattella, \emph{{An Introduction to Particle
  Dark Matter}}, \href{https://doi.org/10.3390/universe5100213}{\emph{Universe}
  {\bfseries 5} (2019) 213} [\href{https://arxiv.org/abs/1910.05610}{{\ttfamily
  1910.05610}}].

\bibitem{Tuominen:2021wrl}
K.~Tuominen, \emph{{Cold Particle Dark Matter}},
  \href{https://doi.org/10.3390/sym13101945}{\emph{Symmetry} {\bfseries 13}
  (2021) 1945}.

\bibitem{Arbey:2021gdg}
A.~Arbey and F.~Mahmoudi, \emph{{Dark matter and the early Universe: a
  review}}, \href{https://doi.org/10.1016/j.ppnp.2021.103865}{\emph{Prog. Part.
  Nucl. Phys.} {\bfseries 119} (2021) 103865}
  [\href{https://arxiv.org/abs/2104.11488}{{\ttfamily 2104.11488}}].

\bibitem{Steigman:2012nb}
G.~Steigman, B.~Dasgupta and J.F.~Beacom, \emph{{Precise Relic WIMP Abundance
  and its Impact on Searches for Dark Matter Annihilation}},
  \href{https://doi.org/10.1103/PhysRevD.86.023506}{\emph{Phys. Rev. D}
  {\bfseries 86} (2012) 023506}
  [\href{https://arxiv.org/abs/1204.3622}{{\ttfamily 1204.3622}}].

\bibitem{McDonald:2001vt}
J.~McDonald, \emph{{Thermally generated gauge singlet scalars as
  selfinteracting dark matter}},
  \href{https://doi.org/10.1103/PhysRevLett.88.091304}{\emph{Phys. Rev. Lett.}
  {\bfseries 88} (2002) 091304}
  [\href{https://arxiv.org/abs/hep-ph/0106249}{{\ttfamily hep-ph/0106249}}].

\bibitem{Hall:2009bx}
L.J.~Hall, K.~Jedamzik, J.~March-Russell and S.M.~West, \emph{{Freeze-In
  Production of FIMP Dark Matter}},
  \href{https://doi.org/10.1007/JHEP03(2010)080}{\emph{JHEP} {\bfseries 03}
  (2010) 080} [\href{https://arxiv.org/abs/0911.1120}{{\ttfamily 0911.1120}}].

\bibitem{Bernal:2017kxu}
N.~Bernal, M.~Heikinheimo, T.~Tenkanen, K.~Tuominen and V.~Vaskonen, \emph{{The
  Dawn of FIMP Dark Matter: A Review of Models and Constraints}},
  \href{https://doi.org/10.1142/S0217751X1730023X}{\emph{Int. J. Mod. Phys. A}
  {\bfseries 32} (2017) 1730023}
  [\href{https://arxiv.org/abs/1706.07442}{{\ttfamily 1706.07442}}].

\bibitem{Covi:1999ty}
L.~Covi, J.E.~Kim and L.~Roszkowski, \emph{{Axinos as cold dark matter}},
  \href{https://doi.org/10.1103/PhysRevLett.82.4180}{\emph{Phys. Rev. Lett.}
  {\bfseries 82} (1999) 4180}
  [\href{https://arxiv.org/abs/hep-ph/9905212}{{\ttfamily hep-ph/9905212}}].

\bibitem{Feng:2003uy}
J.L.~Feng, A.~Rajaraman and F.~Takayama, \emph{{SuperWIMP dark matter signals
  from the early universe}},
  \href{https://doi.org/10.1103/PhysRevD.68.063504}{\emph{Phys. Rev. D}
  {\bfseries 68} (2003) 063504}
  [\href{https://arxiv.org/abs/hep-ph/0306024}{{\ttfamily hep-ph/0306024}}].

\bibitem{Feng:2003xh}
J.L.~Feng, A.~Rajaraman and F.~Takayama, \emph{{Superweakly interacting massive
  particles}}, \href{https://doi.org/10.1103/PhysRevLett.91.011302}{\emph{Phys.
  Rev. Lett.} {\bfseries 91} (2003) 011302}
  [\href{https://arxiv.org/abs/hep-ph/0302215}{{\ttfamily hep-ph/0302215}}].

\bibitem{Johnson:1972apj}
I.~{Johnson}, W.~N., J.~{Harnden}, F.~R. and R.C.~{Haymes}, \emph{{The Spectrum
  of Low-Energy Gamma Radiation from the Galactic-Center Region.}},
  \href{https://doi.org/10.1086/180878}{\emph{Astrophysical Journal, Letters}
  {\bfseries 172} (1972) L1}.

\bibitem{Purcell:1997apj}
W.R.~Purcell, L.-X.~Cheng, D.D.~Dixon, R.L.~Kinzer, J.D.~Kurfess, M.~Leventhal
  et~al., \emph{{OSSE Mapping of Galactic 511 keV Positron Annihilation Line
  Emission}}, \href{https://doi.org/10.1086/304994}{\emph{The Astrophysical
  Journal} {\bfseries 491} (1997) 725}.

\bibitem{Bouchet:2010dj}
L.~Bouchet, J.-P.~Roques and E.~Jourdain, \emph{{On the morphology of the
  electron-positron annihilation emission as seen by SPI/INTEGRAL}},
  \href{https://doi.org/10.1088/0004-637X/720/2/1772}{\emph{Astrophys. J.}
  {\bfseries 720} (2010) 1772}
  [\href{https://arxiv.org/abs/1007.4753}{{\ttfamily 1007.4753}}].

\bibitem{Bouchet:2011fn}
L.~Bouchet, A.W.~Strong, T.A.~Porter, I.V.~Moskalenko, E.~Jourdain and
  J.-P.~Roques, \emph{{Diffuse emission measurement with INTEGRAL/SPI as
  indirect probe of cosmic-ray electrons and positrons}},
  \href{https://doi.org/10.1088/0004-637X/739/1/29}{\emph{Astrophys. J.}
  {\bfseries 739} (2011) 29} [\href{https://arxiv.org/abs/1107.0200}{{\ttfamily
  1107.0200}}].

\bibitem{Siegert:2019tus}
T.~Siegert, R.M.~Crocker, R.~Diehl, M.G.H.~Krause, F.H.~Panther,
  M.M.M.~Pleintinger et~al., \emph{{Constraints on positron annihilation
  kinematics in the inner Galaxy}},
  \href{https://doi.org/10.1051/0004-6361/201833856}{\emph{Astron. Astrophys.}
  {\bfseries 627} (2019) A126}
  [\href{https://arxiv.org/abs/1906.00498}{{\ttfamily 1906.00498}}].

\bibitem{Prantzos:2010wi}
N.~Prantzos et~al., \emph{{The 511 keV emission from positron annihilation in
  the Galaxy}}, \href{https://doi.org/10.1103/RevModPhys.83.1001}{\emph{Rev.
  Mod. Phys.} {\bfseries 83} (2011) 1001}
  [\href{https://arxiv.org/abs/1009.4620}{{\ttfamily 1009.4620}}].

\bibitem{Leane:2022bfm}
R.K.~Leane et~al., \emph{{Snowmass2021 Cosmic Frontier White Paper: Puzzling
  Excesses in Dark Matter Searches and How to Resolve Them}},
  \href{https://arxiv.org/abs/2203.06859}{{\ttfamily 2203.06859}}.

\bibitem{Siegert:2023wus}
T.~Siegert, \emph{{The Positron Puzzle}},  3, 2023,
  \href{https://doi.org/10.1007/s10509-023-04184-4}{DOI}
  [\href{https://arxiv.org/abs/2303.15582}{{\ttfamily 2303.15582}}].

\bibitem{Boehm:2003bt}
C.~Boehm, D.~Hooper, J.~Silk, M.~Casse and J.~Paul, \emph{{MeV dark matter: Has
  it been detected?}},
  \href{https://doi.org/10.1103/PhysRevLett.92.101301}{\emph{Phys. Rev. Lett.}
  {\bfseries 92} (2004) 101301}
  [\href{https://arxiv.org/abs/astro-ph/0309686}{{\ttfamily
  astro-ph/0309686}}].

\bibitem{Hooper:2004qf}
D.~Hooper and L.-T.~Wang, \emph{{Possible evidence for axino dark matter in the
  galactic bulge}},
  \href{https://doi.org/10.1103/PhysRevD.70.063506}{\emph{Phys. Rev. D}
  {\bfseries 70} (2004) 063506}
  [\href{https://arxiv.org/abs/hep-ph/0402220}{{\ttfamily hep-ph/0402220}}].

\bibitem{Picciotto:2004rp}
C.~Picciotto and M.~Pospelov, \emph{{Unstable relics as a source of galactic
  positrons}},
  \href{https://doi.org/10.1016/j.physletb.2004.11.025}{\emph{Phys. Lett. B}
  {\bfseries 605} (2005) 15}
  [\href{https://arxiv.org/abs/hep-ph/0402178}{{\ttfamily hep-ph/0402178}}].

\bibitem{Gunion:2005rw}
J.F.~Gunion, D.~Hooper and B.~McElrath, \emph{{Light neutralino dark matter in
  the NMSSM}}, \href{https://doi.org/10.1103/PhysRevD.73.015011}{\emph{Phys.
  Rev. D} {\bfseries 73} (2006) 015011}
  [\href{https://arxiv.org/abs/hep-ph/0509024}{{\ttfamily hep-ph/0509024}}].

\bibitem{Takahashi:2005kp}
F.~Takahashi and T.T.~Yanagida, \emph{{Unification of dark energy and dark
  matter}}, \href{https://doi.org/10.1016/j.physletb.2006.02.026}{\emph{Phys.
  Lett. B} {\bfseries 635} (2006) 57}
  [\href{https://arxiv.org/abs/hep-ph/0512296}{{\ttfamily hep-ph/0512296}}].

\bibitem{Finkbeiner:2007kk}
D.P.~Finkbeiner and N.~Weiner, \emph{{Exciting Dark Matter and the INTEGRAL/SPI
  511 keV signal}},
  \href{https://doi.org/10.1103/PhysRevD.76.083519}{\emph{Phys. Rev. D}
  {\bfseries 76} (2007) 083519}
  [\href{https://arxiv.org/abs/astro-ph/0702587}{{\ttfamily
  astro-ph/0702587}}].

\bibitem{Huh:2007zw}
J.-H.~Huh, J.E.~Kim, J.-C.~Park and S.C.~Park, \emph{{Galactic 511 keV line
  from MeV milli-charged dark matter}},
  \href{https://doi.org/10.1103/PhysRevD.77.123503}{\emph{Phys. Rev. D}
  {\bfseries 77} (2008) 123503}
  [\href{https://arxiv.org/abs/0711.3528}{{\ttfamily 0711.3528}}].

\bibitem{Pospelov:2007xh}
M.~Pospelov and A.~Ritz, \emph{{The galactic 511 keV line from electroweak
  scale WIMPs}},
  \href{https://doi.org/10.1016/j.physletb.2007.06.027}{\emph{Phys. Lett. B}
  {\bfseries 651} (2007) 208}
  [\href{https://arxiv.org/abs/hep-ph/0703128}{{\ttfamily hep-ph/0703128}}].

\bibitem{Cembranos:2008bw}
J.A.R.~Cembranos and L.E.~Strigari, \emph{{Diffuse MeV Gamma-rays and Galactic
  511 keV Line from Decaying WIMP Dark Matter}},
  \href{https://doi.org/10.1103/PhysRevD.77.123519}{\emph{Phys. Rev. D}
  {\bfseries 77} (2008) 123519}
  [\href{https://arxiv.org/abs/0801.0630}{{\ttfamily 0801.0630}}].

\bibitem{Arkani-Hamed:2008hhe}
N.~Arkani-Hamed, D.P.~Finkbeiner, T.R.~Slatyer and N.~Weiner, \emph{{A Theory
  of Dark Matter}},
  \href{https://doi.org/10.1103/PhysRevD.79.015014}{\emph{Phys. Rev. D}
  {\bfseries 79} (2009) 015014}
  [\href{https://arxiv.org/abs/0810.0713}{{\ttfamily 0810.0713}}].

\bibitem{Khlopov:2008ki}
M.Y.~Khlopov, \emph{{Composite dark matter from stable charged constituents}},
  in \emph{{20th Rencontres de Blois on Challenges in Particle Astrophysics}},
  pp.~345--350, 2008 [\href{https://arxiv.org/abs/0806.3581}{{\ttfamily
  0806.3581}}].

\bibitem{Khlopov:2009hi}
M.Y.~Khlopov, \emph{{The puzzles of dark matter searches}},
  \href{https://doi.org/10.1063/1.3462660}{\emph{AIP Conf. Proc.} {\bfseries
  1241} (2010) 388} [\href{https://arxiv.org/abs/0911.5685}{{\ttfamily
  0911.5685}}].

\bibitem{Chen:2009av}
F.~Chen, J.M.~Cline, A.~Fradette, A.R.~Frey and C.~Rabideau, \emph{{Exciting
  dark matter in the galactic center}},
  \href{https://doi.org/10.1103/PhysRevD.81.043523}{\emph{Phys. Rev. D}
  {\bfseries 81} (2010) 043523}
  [\href{https://arxiv.org/abs/0911.2222}{{\ttfamily 0911.2222}}].

\bibitem{Chen:2009dm}
F.~Chen, J.M.~Cline and A.R.~Frey, \emph{{A New twist on excited dark matter:
  Implications for INTEGRAL, PAMELA/ATIC/PPB-BETS, DAMA}},
  \href{https://doi.org/10.1103/PhysRevD.79.063530}{\emph{Phys. Rev. D}
  {\bfseries 79} (2009) 063530}
  [\href{https://arxiv.org/abs/0901.4327}{{\ttfamily 0901.4327}}].

\bibitem{Finkbeiner:2009mi}
D.P.~Finkbeiner, T.R.~Slatyer, N.~Weiner and I.~Yavin, \emph{{PAMELA, DAMA,
  INTEGRAL and Signatures of Metastable Excited WIMPs}},
  \href{https://doi.org/10.1088/1475-7516/2009/09/037}{\emph{JCAP} {\bfseries
  09} (2009) 037} [\href{https://arxiv.org/abs/0903.1037}{{\ttfamily
  0903.1037}}].

\bibitem{Khlopov:2010pq}
M.Y.~Khlopov, A.G.~Mayorov and E.Y.~Soldatov, \emph{{Composite Dark Matter and
  Puzzles of Dark Matter Searches}},
  \href{https://doi.org/10.1142/S0218271810017962}{\emph{Int. J. Mod. Phys. D}
  {\bfseries 19} (2010) 1385}
  [\href{https://arxiv.org/abs/1003.1144}{{\ttfamily 1003.1144}}].

\bibitem{Cline:2010kv}
J.M.~Cline, A.R.~Frey and F.~Chen, \emph{{Metastable dark matter mechanisms for
  INTEGRAL 511 keV $\gamma$ rays and DAMA/CoGeNT events}},
  \href{https://doi.org/10.1103/PhysRevD.83.083511}{\emph{Phys. Rev. D}
  {\bfseries 83} (2011) 083511}
  [\href{https://arxiv.org/abs/1008.1784}{{\ttfamily 1008.1784}}].

\bibitem{Cline:2012yx}
J.M.~Cline and A.R.~Frey, \emph{{Abelian dark matter models for 511 keV gamma
  rays and direct detection}},
  \href{https://doi.org/10.1002/andp.201200082}{\emph{Annalen Phys.} {\bfseries
  524} (2012) 579} [\href{https://arxiv.org/abs/1204.1965}{{\ttfamily
  1204.1965}}].

\bibitem{Boubekeur:2012eq}
L.~Boubekeur, S.~Dodelson and O.~Vives, \emph{{Cold Positrons from Decaying
  Dark Matter}}, \href{https://doi.org/10.1103/PhysRevD.86.103520}{\emph{Phys.
  Rev. D} {\bfseries 86} (2012) 103520}
  [\href{https://arxiv.org/abs/1206.3076}{{\ttfamily 1206.3076}}].

\bibitem{Cudell:2014jba}
J.-R.~Cudell, M.Y.~Khlopov and Q.~Wallemacq, \emph{{Dark atoms and the
  positron-annihilation-line excess in the galactic bulge}},
  \href{https://arxiv.org/abs/1401.5228}{{\ttfamily 1401.5228}}.

\bibitem{Cudell:2014wca}
J.-R.~Cudell, M.Y.~Khlopov and Q.~Wallemacq, \emph{{Effects of dark atom
  excitations}}, \href{https://doi.org/10.1142/S0217732314400069}{\emph{Mod.
  Phys. Lett. A} {\bfseries 29} (2014) 1440006}
  [\href{https://arxiv.org/abs/1411.1655}{{\ttfamily 1411.1655}}].

\bibitem{Farzan:2017hol}
Y.~Farzan and M.~Rajaee, \emph{{Pico-charged intermediate particles rescue dark
  matter interpretation of 511 keV signal}},
  \href{https://doi.org/10.1007/JHEP12(2017)083}{\emph{JHEP} {\bfseries 12}
  (2017) 083} [\href{https://arxiv.org/abs/1708.01137}{{\ttfamily
  1708.01137}}].

\bibitem{Jia:2017iyc}
L.-B.~Jia, \emph{{Explanation of the 511 keV line: cascade annihilating dark
  matter with the$^8$ Be anomaly}},
  \href{https://doi.org/10.1140/epjc/s10052-018-5555-0}{\emph{Eur. Phys. J. C}
  {\bfseries 78} (2018) 112}
  [\href{https://arxiv.org/abs/1710.03906}{{\ttfamily 1710.03906}}].

\bibitem{Cai:2020fnq}
R.-G.~Cai, Y.-C.~Ding, X.-Y.~Yang and Y.-F.~Zhou, \emph{{Constraints on a mixed
  model of dark matter particles and primordial black holes from the galactic
  511 keV line}},
  \href{https://doi.org/10.1088/1475-7516/2021/03/057}{\emph{JCAP} {\bfseries
  03} (2021) 057} [\href{https://arxiv.org/abs/2007.11804}{{\ttfamily
  2007.11804}}].

\bibitem{Ema:2020fit}
Y.~Ema, F.~Sala and R.~Sato, \emph{{Dark matter models for the 511 keV galactic
  line predict keV electron recoils on Earth}},
  \href{https://doi.org/10.1140/epjc/s10052-021-08899-y}{\emph{Eur. Phys. J. C}
  {\bfseries 81} (2021) 129}
  [\href{https://arxiv.org/abs/2007.09105}{{\ttfamily 2007.09105}}].

\bibitem{Lin:2022mqe}
W.~Lin and T.T.~Yanagida, \emph{{Confronting the Galactic 511~keV emission with
  B-L gauge boson dark matter}},
  \href{https://doi.org/10.1103/PhysRevD.106.075012}{\emph{Phys. Rev. D}
  {\bfseries 106} (2022) 075012}
  [\href{https://arxiv.org/abs/2205.08171}{{\ttfamily 2205.08171}}].

\bibitem{Sheng:2023iup}
J.~Sheng, Y.~Cheng, W.~Lin and T.T.~Yanagida, \emph{{F\'eeton (B-L gauge boson)
  dark matter for the 511-keV gamma-ray excess and the prediction of low-energy
  neutrino flux*}}, \href{https://doi.org/10.1088/1674-1137/ad4af3}{\emph{Chin.
  Phys. C} {\bfseries 48} (2024) 083104}
  [\href{https://arxiv.org/abs/2310.05420}{{\ttfamily 2310.05420}}].

\bibitem{Feng:2024nkh}
W.-Z.~Feng and Z.-H.~Zhang, \emph{{Freeze-in Dark Matter Explanation of the
  Galactic 511 keV Signal}},
  \href{https://arxiv.org/abs/2405.19431}{{\ttfamily 2405.19431}}.

\bibitem{Dolgov:2023sst}
A.D.~Dolgov and A.S.~Rudenko, \emph{{Conversion of Protons to Positrons by a
  Black Hole}}, \href{https://doi.org/10.1134/S1547477124700821}{\emph{Phys.
  Part. Nucl. Lett.} {\bfseries 21} (2024) 865}
  [\href{https://arxiv.org/abs/2308.01689}{{\ttfamily 2308.01689}}].

\bibitem{Beacom:2005qv}
J.F.~Beacom and H.~Yuksel, \emph{{Stringent constraint on galactic positron
  production}},
  \href{https://doi.org/10.1103/PhysRevLett.97.071102}{\emph{Phys. Rev. Lett.}
  {\bfseries 97} (2006) 071102}
  [\href{https://arxiv.org/abs/astro-ph/0512411}{{\ttfamily
  astro-ph/0512411}}].

\bibitem{Ascasibar:2005rw}
Y.~Ascasibar, P.~Jean, C.~Boehm and J.~Knoedlseder, \emph{{Constraints on dark
  matter and the shape of the Milky Way dark halo from the 511-keV line}},
  \href{https://doi.org/10.1111/j.1365-2966.2006.10226.x}{\emph{Mon. Not. Roy.
  Astron. Soc.} {\bfseries 368} (2006) 1695}
  [\href{https://arxiv.org/abs/astro-ph/0507142}{{\ttfamily
  astro-ph/0507142}}].

\bibitem{Vincent:2012an}
A.C.~Vincent, P.~Martin and J.M.~Cline, \emph{{Interacting dark matter
  contribution to the Galactic 511 keV gamma ray emission: constraining the
  morphology with INTEGRAL/SPI observations}},
  \href{https://doi.org/10.1088/1475-7516/2012/04/022}{\emph{JCAP} {\bfseries
  04} (2012) 022} [\href{https://arxiv.org/abs/1201.0997}{{\ttfamily
  1201.0997}}].

\bibitem{Wilkinson:2016gsy}
R.J.~Wilkinson, A.C.~Vincent, C.~B\oe{}hm and C.~McCabe, \emph{{Ruling out the
  light weakly interacting massive particle explanation of the Galactic 511 keV
  line}}, \href{https://doi.org/10.1103/PhysRevD.94.103525}{\emph{Phys. Rev. D}
  {\bfseries 94} (2016) 103525}
  [\href{https://arxiv.org/abs/1602.01114}{{\ttfamily 1602.01114}}].

\bibitem{Cappiello:2023qwl}
C.V.~Cappiello, M.~Jafs and A.C.~Vincent, \emph{{The morphology of exciting
  dark matter and the galactic 511 keV signal}},
  \href{https://doi.org/10.1088/1475-7516/2023/11/003}{\emph{JCAP} {\bfseries
  11} (2023) 003} [\href{https://arxiv.org/abs/2307.15114}{{\ttfamily
  2307.15114}}].

\bibitem{DelaTorreLuque:2023cef}
P.~De~la Torre~Luque, S.~Balaji and J.~Silk, \emph{{New 511 keV line data
  provides strongest sub-GeV dark matter constraints}},
  \href{https://arxiv.org/abs/2312.04907}{{\ttfamily 2312.04907}}.

\bibitem{Babu:1997st}
K.S.~Babu, C.F.~Kolda and J.~March-Russell, \emph{{Implications of generalized
  Z - Z-prime mixing}},
  \href{https://doi.org/10.1103/PhysRevD.57.6788}{\emph{Phys. Rev. D}
  {\bfseries 57} (1998) 6788}
  [\href{https://arxiv.org/abs/hep-ph/9710441}{{\ttfamily hep-ph/9710441}}].

\bibitem{Choi:2021yps}
S.-M.~Choi, J.~Kim, P.~Ko and J.~Li, \emph{{A multi-component SIMP model with
  U(1)$_{X}$\textrightarrow{} Z$_{2}$ \texttimes{} Z$_{3}$}},
  \href{https://doi.org/10.1007/JHEP09(2021)028}{\emph{JHEP} {\bfseries 09}
  (2021) 028} [\href{https://arxiv.org/abs/2103.05956}{{\ttfamily
  2103.05956}}].

\bibitem{Bauer:2018egk}
M.~Bauer, S.~Diefenbacher, T.~Plehn, M.~Russell and D.A.~Camargo, \emph{{Dark
  Matter in Anomaly-Free Gauge Extensions}},
  \href{https://doi.org/10.21468/SciPostPhys.5.4.036}{\emph{SciPost Phys.}
  {\bfseries 5} (2018) 036} [\href{https://arxiv.org/abs/1805.01904}{{\ttfamily
  1805.01904}}].

\bibitem{Fabbrichesi:2020wbt}
M.~Fabbrichesi, E.~Gabrielli and G.~Lanfranchi, \emph{{The Dark Photon}},
  \href{https://arxiv.org/abs/2005.01515}{{\ttfamily 2005.01515}}.

\bibitem{Costa:2022lpy}
F.~Costa, S.~Khan and J.~Kim, \emph{{A two-component vector WIMP \textemdash{}
  fermion FIMP dark matter model with an extended seesaw mechanism}},
  \href{https://doi.org/10.1007/JHEP12(2022)165}{\emph{JHEP} {\bfseries 12}
  (2022) 165} [\href{https://arxiv.org/abs/2209.13653}{{\ttfamily
  2209.13653}}].

\bibitem{Fermi-LAT:2015kyq}
{\scshape Fermi-LAT} collaboration, \emph{{Updated search for spectral lines
  from Galactic dark matter interactions with pass 8 data from the Fermi Large
  Area Telescope}},
  \href{https://doi.org/10.1103/PhysRevD.91.122002}{\emph{Phys. Rev. D}
  {\bfseries 91} (2015) 122002}
  [\href{https://arxiv.org/abs/1506.00013}{{\ttfamily 1506.00013}}].

\bibitem{Navarro:1995iw}
J.F.~Navarro, C.S.~Frenk and S.D.M.~White, \emph{{The Structure of cold dark
  matter halos}}, \href{https://doi.org/10.1086/177173}{\emph{Astrophys. J.}
  {\bfseries 462} (1996) 563}
  [\href{https://arxiv.org/abs/astro-ph/9508025}{{\ttfamily
  astro-ph/9508025}}].

\bibitem{Cirelli:2010xx}
M.~Cirelli, G.~Corcella, A.~Hektor, G.~Hutsi, M.~Kadastik, P.~Panci et~al.,
  \emph{{PPPC 4 DM ID: A Poor Particle Physicist Cookbook for Dark Matter
  Indirect Detection}},
  \href{https://doi.org/10.1088/1475-7516/2012/10/E01}{\emph{JCAP} {\bfseries
  03} (2011) 051} [\href{https://arxiv.org/abs/1012.4515}{{\ttfamily
  1012.4515}}].

\bibitem{Siegert:2015knp}
T.~Siegert, R.~Diehl, G.~Khachatryan, M.G.H.~Krause, F.~Guglielmetti,
  J.~Greiner et~al., \emph{{Gamma-ray spectroscopy of Positron Annihilation in
  the Milky Way}},
  \href{https://doi.org/10.1051/0004-6361/201527510}{\emph{Astron. Astrophys.}
  {\bfseries 586} (2016) A84}
  [\href{https://arxiv.org/abs/1512.00325}{{\ttfamily 1512.00325}}].

\bibitem{Pospelov:2007mp}
M.~Pospelov, A.~Ritz and M.B.~Voloshin, \emph{{Secluded WIMP Dark Matter}},
  \href{https://doi.org/10.1016/j.physletb.2008.02.052}{\emph{Phys. Lett. B}
  {\bfseries 662} (2008) 53} [\href{https://arxiv.org/abs/0711.4866}{{\ttfamily
  0711.4866}}].

\bibitem{Khan:2023uii}
S.~Khan, J.~Kim and P.~Ko, \emph{{Interplay between Higgs inflation and dark
  matter models with dark U(1) gauge symmetry}},
  \href{https://doi.org/10.1007/JHEP05(2024)250}{\emph{JHEP} {\bfseries 05}
  (2024) 250} [\href{https://arxiv.org/abs/2309.07839}{{\ttfamily
  2309.07839}}].

\bibitem{Alguero:2023zol}
G.~Alguero, G.~Belanger, F.~Boudjema, S.~Chakraborti, A.~Goudelis, S.~Kraml
  et~al., \emph{{micrOMEGAs 6.0: N-component dark matter}},
  \href{https://doi.org/10.1016/j.cpc.2024.109133}{\emph{Comput. Phys. Commun.}
  {\bfseries 299} (2024) 109133}
  [\href{https://arxiv.org/abs/2312.14894}{{\ttfamily 2312.14894}}].

\bibitem{Gondolo:1990dk}
P.~Gondolo and G.~Gelmini, \emph{{Cosmic abundances of stable particles:
  Improved analysis}},
  \href{https://doi.org/10.1016/0550-3213(91)90438-4}{\emph{Nucl. Phys. B}
  {\bfseries 360} (1991) 145}.

\bibitem{Edsjo:1997bg}
J.~Edsjo and P.~Gondolo, \emph{{Neutralino relic density including
  coannihilations}},
  \href{https://doi.org/10.1103/PhysRevD.56.1879}{\emph{Phys. Rev. D}
  {\bfseries 56} (1997) 1879}
  [\href{https://arxiv.org/abs/hep-ph/9704361}{{\ttfamily hep-ph/9704361}}].

\bibitem{CMS:2022qva}
{\scshape CMS} collaboration, \emph{{Search for invisible decays of the Higgs
  boson produced via vector boson fusion in proton-proton collisions at
  s=13\,\,TeV}}, \href{https://doi.org/10.1103/PhysRevD.105.092007}{\emph{Phys.
  Rev. D} {\bfseries 105} (2022) 092007}
  [\href{https://arxiv.org/abs/2201.11585}{{\ttfamily 2201.11585}}].

\bibitem{Junnarkar:2013ac}
P.~Junnarkar and A.~Walker-Loud, \emph{{Scalar strange content of the nucleon
  from lattice QCD}},
  \href{https://doi.org/10.1103/PhysRevD.87.114510}{\emph{Phys. Rev. D}
  {\bfseries 87} (2013) 114510}
  [\href{https://arxiv.org/abs/1301.1114}{{\ttfamily 1301.1114}}].

\bibitem{XENON:2023cxc}
{\scshape XENON} collaboration, \emph{{First Dark Matter Search with Nuclear
  Recoils from the XENONnT Experiment}},
  \href{https://doi.org/10.1103/PhysRevLett.131.041003}{\emph{Phys. Rev. Lett.}
  {\bfseries 131} (2023) 041003}
  [\href{https://arxiv.org/abs/2303.14729}{{\ttfamily 2303.14729}}].

\bibitem{Chu:2023jyb}
X.~Chu and J.~Pradler, \emph{{Minimal mass of thermal dark matter and the
  viability of millicharged particles affecting 21-cm cosmology}},
  \href{https://doi.org/10.1103/PhysRevD.109.103510}{\emph{Phys. Rev. D}
  {\bfseries 109} (2024) 103510}
  [\href{https://arxiv.org/abs/2310.06611}{{\ttfamily 2310.06611}}].

\bibitem{Sabti:2019mhn}
N.~Sabti, J.~Alvey, M.~Escudero, M.~Fairbairn and D.~Blas, \emph{{Refined
  Bounds on MeV-scale Thermal Dark Sectors from BBN and the CMB}},
  \href{https://doi.org/10.1088/1475-7516/2020/01/004}{\emph{JCAP} {\bfseries
  01} (2020) 004} [\href{https://arxiv.org/abs/1910.01649}{{\ttfamily
  1910.01649}}].

\bibitem{EscuderoAbenza:2020cmq}
M.~Escudero~Abenza, \emph{{Precision early universe thermodynamics made simple:
  $N_{\rm eff}$ and neutrino decoupling in the Standard Model and beyond}},
  \href{https://doi.org/10.1088/1475-7516/2020/05/048}{\emph{JCAP} {\bfseries
  05} (2020) 048} [\href{https://arxiv.org/abs/2001.04466}{{\ttfamily
  2001.04466}}].

\bibitem{Akita:2020szl}
K.~Akita and M.~Yamaguchi, \emph{{A precision calculation of relic neutrino
  decoupling}},
  \href{https://doi.org/10.1088/1475-7516/2020/08/012}{\emph{JCAP} {\bfseries
  08} (2020) 012} [\href{https://arxiv.org/abs/2005.07047}{{\ttfamily
  2005.07047}}].

\bibitem{Bennett:2020zkv}
J.J.~Bennett, G.~Buldgen, P.F.~De~Salas, M.~Drewes, S.~Gariazzo, S.~Pastor
  et~al., \emph{{Towards a precision calculation of $N_{\rm eff}$ in the
  Standard Model II: Neutrino decoupling in the presence of flavour
  oscillations and finite-temperature QED}},
  \href{https://doi.org/10.1088/1475-7516/2021/04/073}{\emph{JCAP} {\bfseries
  04} (2021) 073} [\href{https://arxiv.org/abs/2012.02726}{{\ttfamily
  2012.02726}}].

\bibitem{Aloni:2023tff}
D.~Aloni, M.~Joseph, M.~Schmaltz and N.~Weiner, \emph{{Dark Radiation from
  Neutrino Mixing after Big Bang Nucleosynthesis}},
  \href{https://doi.org/10.1103/PhysRevLett.131.221001}{\emph{Phys. Rev. Lett.}
  {\bfseries 131} (2023) 221001}
  [\href{https://arxiv.org/abs/2301.10792}{{\ttfamily 2301.10792}}].

\bibitem{Slatyer:2015jla}
T.R.~Slatyer, \emph{{Indirect dark matter signatures in the cosmic dark ages.
  I. Generalizing the bound on s-wave dark matter annihilation from Planck
  results}}, \href{https://doi.org/10.1103/PhysRevD.93.023527}{\emph{Phys. Rev.
  D} {\bfseries 93} (2016) 023527}
  [\href{https://arxiv.org/abs/1506.03811}{{\ttfamily 1506.03811}}].

\bibitem{Cosme:2023xpa}
C.~Cosme, F.~Costa and O.~Lebedev, \emph{{Freeze-in at stronger coupling}},
  \href{https://doi.org/10.1103/PhysRevD.109.075038}{\emph{Phys. Rev. D}
  {\bfseries 109} (2024) 075038}
  [\href{https://arxiv.org/abs/2306.13061}{{\ttfamily 2306.13061}}].

\bibitem{Silva-Malpartida:2023yks}
J.~Silva-Malpartida, N.~Bernal, J.~Jones-P\'erez and R.A.~Lineros, \emph{{From
  WIMPs to FIMPs with low~reheating~temperatures}},
  \href{https://doi.org/10.1088/1475-7516/2023/09/015}{\emph{JCAP} {\bfseries
  09} (2023) 015} [\href{https://arxiv.org/abs/2306.14943}{{\ttfamily
  2306.14943}}].

\bibitem{Ahmed:2017dbb}
A.~Ahmed, M.~Duch, B.~Grzadkowski and M.~Iglicki, \emph{{Multi-Component Dark
  Matter: the vector and fermion case}},
  \href{https://doi.org/10.1140/epjc/s10052-018-6371-2}{\emph{Eur. Phys. J. C}
  {\bfseries 78} (2018) 905}
  [\href{https://arxiv.org/abs/1710.01853}{{\ttfamily 1710.01853}}].

\bibitem{Konig:2016dzg}
J.~K\"onig, A.~Merle and M.~Totzauer, \emph{{keV Sterile Neutrino Dark Matter
  from Singlet Scalar Decays: The Most General Case}},
  \href{https://doi.org/10.1088/1475-7516/2016/11/038}{\emph{JCAP} {\bfseries
  11} (2016) 038} [\href{https://arxiv.org/abs/1609.01289}{{\ttfamily
  1609.01289}}].

\bibitem{Biswas:2017ait}
A.~Biswas, S.~Choubey, L.~Covi and S.~Khan, \emph{{Explaining the 3.5 keV X-ray
  Line in a ${L_{\mu}-L_{\tau}}$ Extension of the Inert Doublet Model}},
  \href{https://doi.org/10.1088/1475-7516/2018/02/002}{\emph{JCAP} {\bfseries
  02} (2018) 002} [\href{https://arxiv.org/abs/1711.00553}{{\ttfamily
  1711.00553}}].

\bibitem{Decant:2021mhj}
Q.~Decant, J.~Heisig, D.C.~Hooper and L.~Lopez-Honorez,
  \emph{{Lyman-\ensuremath{\alpha} constraints on freeze-in and superWIMPs}},
  \href{https://doi.org/10.1088/1475-7516/2022/03/041}{\emph{JCAP} {\bfseries
  03} (2022) 041} [\href{https://arxiv.org/abs/2111.09321}{{\ttfamily
  2111.09321}}].

\bibitem{Khan:2025yko}
S.~Khan and H.M.~Lee, \emph{{WIMP-FIMP option and neutrino masses via a novel
  anomaly-free B-L symmetry}},
  \href{https://arxiv.org/abs/2503.02635}{{\ttfamily 2503.02635}}.

\bibitem{Covi:2022hqb}
L.~Covi and S.~Khan, \emph{{Axion and FIMP dark matter in a
  \ensuremath{\mathsf{U}}(1) extension of the Standard Model}},
  \href{https://doi.org/10.1088/1475-7516/2022/09/064}{\emph{JCAP} {\bfseries
  09} (2022) 064} [\href{https://arxiv.org/abs/2205.10150}{{\ttfamily
  2205.10150}}].

\bibitem{Abdallah:2019svm}
W.~Abdallah, S.~Choubey and S.~Khan, \emph{{FIMP dark matter candidate(s) in a
  $B - L$ model with inverse seesaw mechanism}},
  \href{https://doi.org/10.1007/JHEP06(2019)095}{\emph{JHEP} {\bfseries 06}
  (2019) 095} [\href{https://arxiv.org/abs/1904.10015}{{\ttfamily
  1904.10015}}].

\bibitem{Covi:2002vw}
L.~Covi, L.~Roszkowski and M.~Small, \emph{{Effects of squark processes on the
  axino CDM abundance}},
  \href{https://doi.org/10.1088/1126-6708/2002/07/023}{\emph{JHEP} {\bfseries
  07} (2002) 023} [\href{https://arxiv.org/abs/hep-ph/0206119}{{\ttfamily
  hep-ph/0206119}}].

\end{thebibliography}\endgroup

\end{document}